\documentclass[aps,prb,groupedaddress,superscriptaddress,twocolumn]{revtex4}

\usepackage{ae,aecompl}
\usepackage{amssymb}
\usepackage{helvet}

\usepackage[T1]{fontenc}
\usepackage[latin9]{inputenc}
\usepackage{graphicx}
\usepackage{amsmath}
\usepackage{epstopdf}
\usepackage{esint}
\usepackage{color}
\usepackage{ulem}
\usepackage{setspace}

\newcommand{\vect}[1]{\boldsymbol{#1}}

\begin{document}

\title{Quantum meets classical phase transition: Low-temperature anomaly in disordered superconductors near $B_{c2}$}

\author{Benjamin Sac\'ep\'e}
\affiliation{Univ. Grenoble Alpes, CNRS, Grenoble INP, Institut N\'{e}el, 38000 Grenoble, France}
\author{Johanna Seidemann}
\affiliation{Univ. Grenoble Alpes, CNRS, Grenoble INP, Institut N\'{e}el, 38000 Grenoble, France}
\author{Fr\'ed\'eric Gay}
\affiliation{Univ. Grenoble Alpes, CNRS, Grenoble INP, Institut N\'{e}el, 38000 Grenoble, France}
\author{Kevin Davenport}
\affiliation{Department of Physics and Astronomy, University of Utah, Salt Lake City, Utah 84112, USA}
\author{Andrey Rogachev}
\affiliation{Department of Physics and Astronomy, University of Utah, Salt Lake City, Utah 84112, USA}
\author{Maoz Ovadia}
\affiliation{Department of Condensed Matter Physics, The Weizmann Institute of Science, Rehovot 76100, Israel.}
\author{Karen Michaeli}
\affiliation{Department of Condensed Matter Physics, The Weizmann Institute of Science, Rehovot 76100, Israel.}
\author{Mikhail V. Feigel'man}
\affiliation{L. D. Landau Institute for Theoretical Physics, Chernogolovka, 142432, Moscow region, Russia}
\affiliation{Skolkovo Institute of Science and Technology, Moscow 143026, Russia}

\begin{abstract}
\textbf{Strongly disordered superconductors in a magnetic field display many characteristic properties of type-II superconductivity--- except at low temperatures where an anomalous linear $T$-dependence of the resistive critical field  $B_{c2}$  is routinely observed. This behavior violates the conventional theory of superconductivity, and its origin remains a long-standing puzzle. Here we report on systematic measurements of the critical magnetic field and current on amorphous indium oxide films of various levels of disorder. Surprisingly, our measurements  show that the $B_{c2}$ anomaly near zero-temperature is accompanied by a clear mean-field like scaling behavior of the critical current.   We demonstrate theoretically that these are consequences of the vortex-glass ground state and its thermal fluctuations. This theory further predicts the linear-$T$ anomaly to occur in films as well as bulk superconductors with a slope that depends on the normal-state sheet resistance---in agreement with experimental data. Thus, our combined experimental and theoretical study reveals  universal low-temperature behavior of $B_{c2}$ in a large class of disordered superconductors. }
\end{abstract}

\maketitle

The magnetic-field tuned transition of disordered superconductors continues to surprise as well as to pose intriguing and challenging puzzles. A wealth of experimental results obtained over decades of study still defies current theoretical understanding. The anomalous temperature  dependence of the resistive critical field $B_{c2}(T)$ near the quantum critical point between superconductor and normal metal is a well known example. Within the conventional Bardeen-Cooper-Schrieffer theory,  $B_{c2}(T)$ is expected to saturate at low temperatures~\cite{class1,class2}. In contrast, a strong upturn of $B_{c2}$ with a \textit{linear}-$T$ dependence as $T\rightarrow0$ has been observed in numerous disordered superconductors. These systems range from alloys  and  oxides, both in thin films~\cite{Tenhover81,upturn1,upturn2,Graybeal84,upturn3,upturn4,SS1} and bulk~\cite{Ren13},  to boron-doped diamond~\cite{Bustarret04} as well as gallium monolayers~\cite{GaChina}. 

Substantial  theoretical efforts~\cite{SpivakZhou,LarkinGalitski,Coffey85,Sadovskii97,Smith00} have been unable to fully resolve   the origin of this anomalous behavior.  The main challenge lies in the complexity of these systems and the subtle interplay between strong fluctuations, disorder and vortex physics. The prevailing explanation for the low-$T$ anomaly of $B_{c2}(T)$ is based on mesoscopic fluctuations~\cite{SpivakZhou,LarkinGalitski}, which result in a spatially inhomogeneous superconducting order parameter.   A recent alternative interpretation invokes a quantum Griffiths singularity to account for the upturn in $B_{c2}(T)$ observed in ultra-thin gallium films~\cite{GaChina}. While these theoretical approaches are generally plausible, they predict an exponential increase of $B_{c2}(T)$ at very low $T$, which manifestly does not capture the specific \textit{linear} dependence measured in disordered superconductors~\cite{Tenhover81,upturn1,upturn2,Graybeal84,upturn3,upturn4,Bustarret04,Ren13,GaChina,SS1}.  
 
To gain new insights on the underlying physical mechanism it is desirable to not only study $B_{c2}(T)$, but to also extract information on additional characteristic  quantities such as the superfluid stiffness.  We therefore  conducted systematic  measurements of both $B_{c2}$ and the critical current $j_{c}$ in films of amorphous indium oxide (a:InO), a prototypical disordered superconductor. In the absence of magnetic field or when vortices are strongly pinned by disorder (i.e., form a vortex glass) the superfluid stiffness can be directly related to the critical current. This is expected to apply to all materials that exhibit the low-temperature $B_{c2}$ anomaly~\cite{Tenhover81,upturn1,upturn2,Graybeal84,upturn3,upturn4,Bustarret04,Ren13,GaChina,SS1}. Consequently, 
measurements of $j_c(B)$ near zero temperature  provide access to the critical behavior of the superfluid density $\rho_s(B)$ near $B_{c2}(0)$, where the low-$T$ anomaly develops. 

The key experimental finding of this work is that the (well-established) linear $T$-dependence of $B_{c2}(T)$ at low temperatures is accompanied by a \textit{power-law} dependence of the critical current  on $B$.  The critical exponent of $j_{c}(B)\sim|B-B_{c2}|^{\upsilon}$ is found to be $ \upsilon \simeq 1.6$.  As explained below, this is consistent with the mean field value $ \upsilon=3/2$ (but not with the mesoscopic fluctuation scenario~\cite{SpivakZhou,LarkinGalitski} which predicts an exponential dependence~\cite{LarkinGalitski}).  Our unexpected finding has direct implication for the critical behavior of $\rho_s(B)$, and demands a revised theory of disordered superconductors in the presence of magnetic field. We therefore complement our experimental work with a comprehensive theoretical study, which identifies the key to understanding the linear  $T$-dependence of $B_{c2}$ in the vortex glass. When vortices are strongly pinned by impurities, their presence only weakly affects the $T=0$ limit of superfluid stiffness and critical current. As a result, both exhibit mean-field-like dependence on the magnetic field.  In contrast, the  temperature variation of the superfluid stiffness is strongly affected  by thermal fluctuations  of the vortex glass.  This gives rise  to the observed linear $T$-dependence of $B_{c2}(T)$ near the quantum critical point. Moreover, we predict a strong dependence of the slope $dB_{c2}(T)/d{T}|_{T\rightarrow0}$ on the sheet resistance, and consequently on the film thickness, which we confirm experimentally.


\textbf{Low-$T$ anomaly near $B_{c2}(0)$}

In this study we focus on a series of a:InO samples, which exhibit a critical temperature between $T_c=3$ K and $3.5$ K and a sheet resistance before transition between $2\, k \Omega $ and $1.2\, k \Omega $ (see Table S1 in SI).  Those samples are far from the disorder-tuned superconductor-insulator transition and behave in many ways as standard dirty superconductors. Their magneto-transport properties upon varying disorder are presented in our previous work~\cite{SS1}. Moreover, to demonstrate the universal role of disorder in the low-$T$ anomaly of $B_{c2}(T)$, we extended our measurements to another thin film material, amorphous molybdenum-germanium~\cite{Kim12} (MoGe); the characterizing parameters as well as the experimental data are presented in the SI.

\begin{figure}[h!]
	\includegraphics[width=1\linewidth]{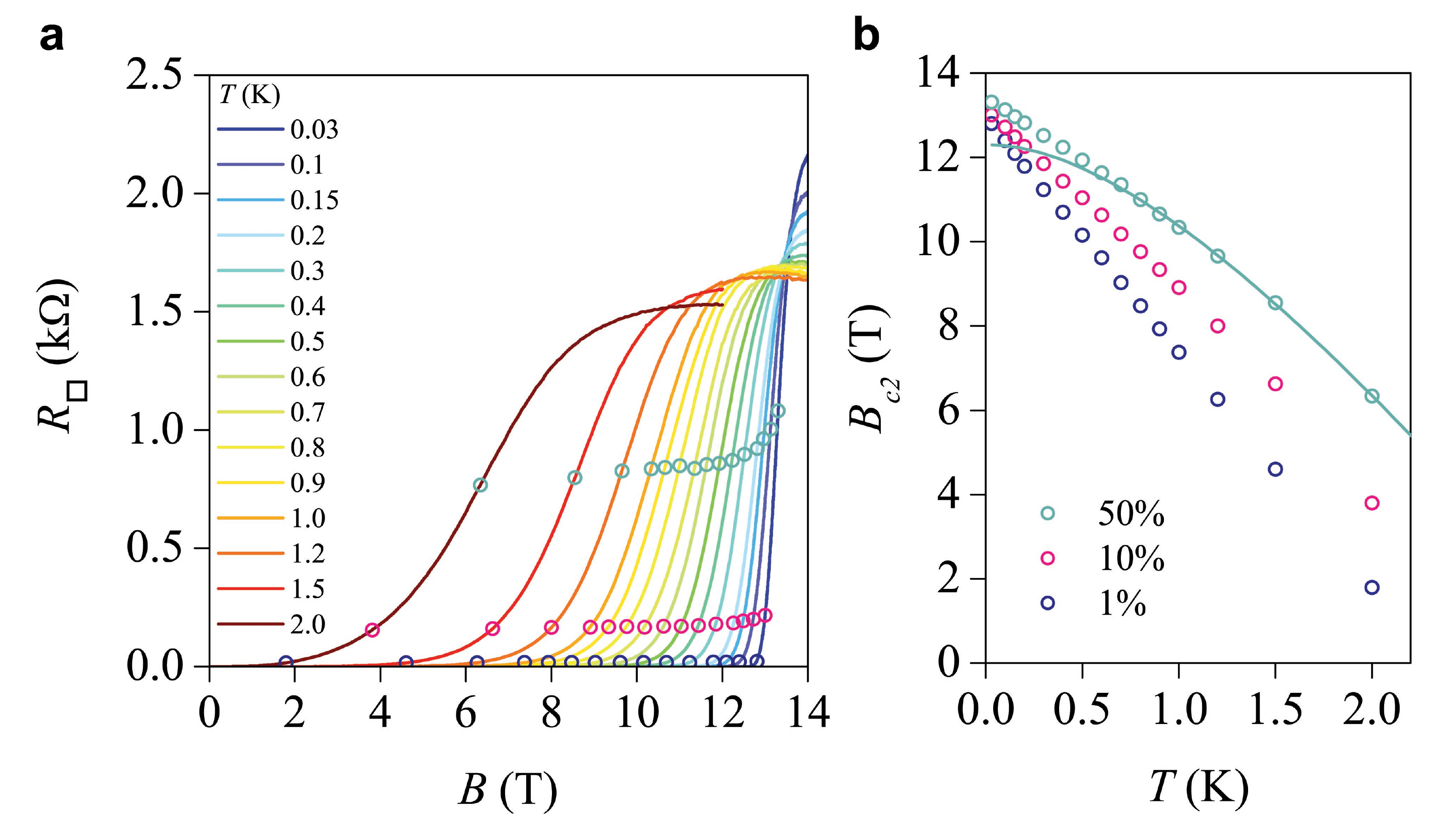}
	\caption{\textbf{Low-$T$ anomaly of the upper critical field $B_{c2}(T)$.} \textbf{a}, $ R_{\square} $ versus $ B $ measured at fixed temperatures for sample J033. Open circles indicate the determination of $B_{c2}(T)$ using three different criteria, namely 50, 10 and 1\% of the high field normal state resistance. \textbf{b}, Extracted $ B_{c2}(T)$ values from the measurement shown in \textbf{a}, plotted versus $ T $. The solid line is a high temperature fit using the theory for dirty superconductors~\cite{class1,class2}.}
	\label{fig1}
\end{figure}


Figure \ref{fig1}a displays the magneto-resistance isotherms of sample J033 measured down to $0.03$ K. 
We define the critical magnetic field $B_{c2}$ through the resistive transition, i.e., the onset of superconducting phase coherence. This critical field does not coincide with the one associated with the pairing instability (see also the discussion in Section IV). To determine $B_{c2}$ at each temperature we used three different criteria, namely $1$, $10$ and $50\%$ of the normal state resistance at high field, indicated by open dots on magneto-resistance isotherms.  The resulting $B_{c2}$ versus $T$ curves are shown in Fig. \ref{fig1}b together with a fit (solid-line) of the high-temperature data with the theory for dirty superconductors~\cite{class1,class2}.  We see that $B_{c2}(T) $ deviates below $\lesssim 1$K from the fit and increases linearly with decreasing $T$ down to our base temperature of 0.03 K (see SI). This deviation, which is the focus of this work, is independent of the criterion used to determine $B_{c2}$. 

\begin{figure}[h!]
	\includegraphics[bb=100 100 1300 690,width=1\linewidth]{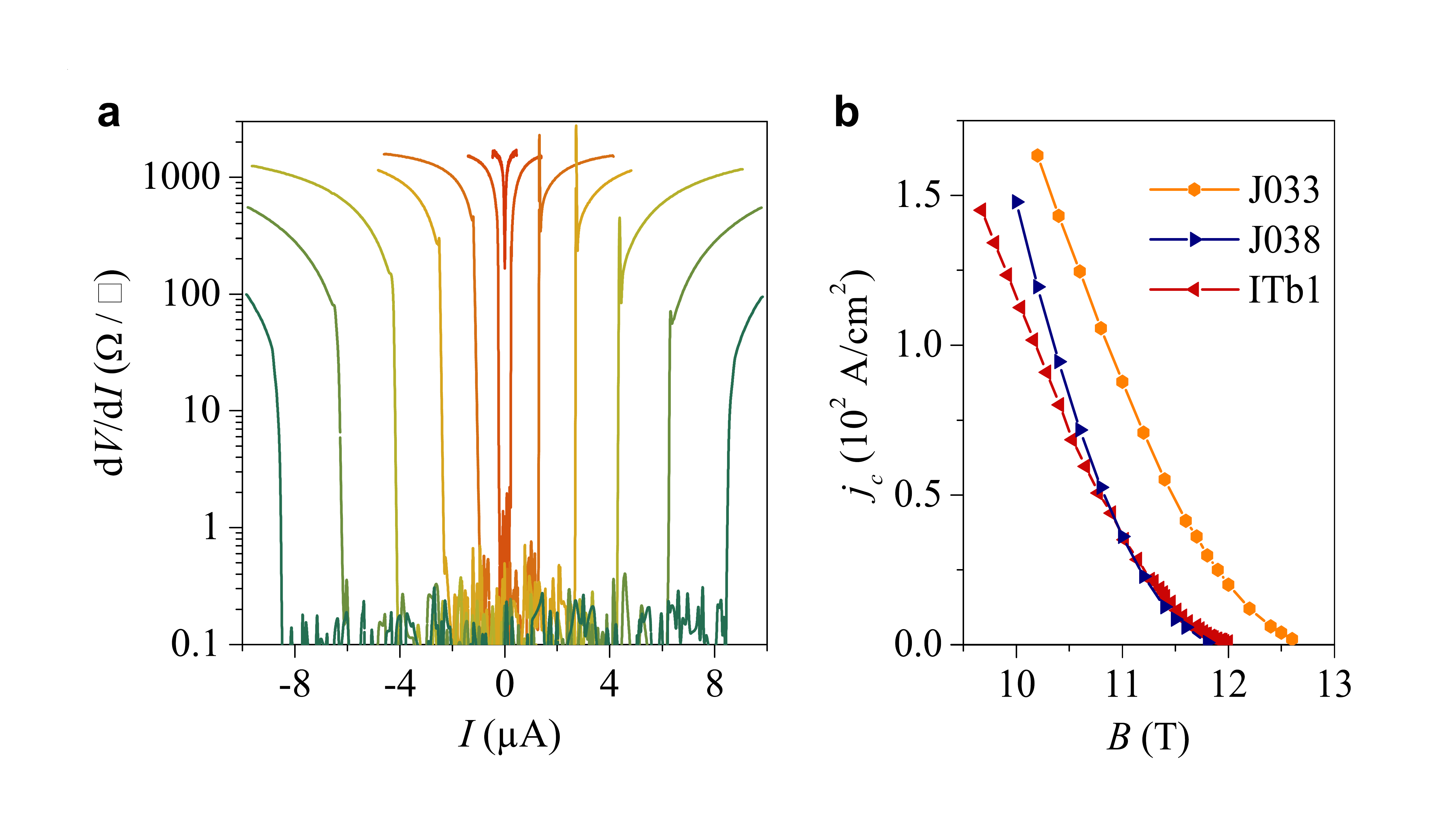}
	\caption{\textbf{Critical current density $ j_c $ near $T=0$ and $B_{c2}(0)$.} \textbf{a}, $ dVdI $ versus $ I $ of sample J033 measured at $T=0.03$ K and at magnetic fields of 10.4, 10.8, 11.2, 11.6, 12, 12.5 and 13 T (green to red curves respectively). \textbf{b}, $ j_c $ versus $ B $ of both samples J033, ITb1 and J038. Each values of $ j_c (B)$ were extracted from $ dVdI $ versus $I$ measurements as shown in \textbf{a} at the resistance value of $ dVdI = 10  \Omega/\square $.}
	\label{fig2}
\end{figure}

An important aspect of the linear dependence of $B_{c2} (T)$ is that it persists down to our lowest $T$ approaching the quantum phase transition QPT at $B_{c2}(0) $. We shall argue below that the linear dependence of $B_{c2} (T)$ can be understood by analyzing the critical behavior of the superfluid density in the vicinity of this QPT.

\textbf{Critical regime of the critical current density}

We turn to the study of the $B$-evolution of the critical current density $j_c$ at our lowest temperature and focus on three samples J033, ITb1 and J038. We systematically measured the differential resistance $dV/dI$ versus current-bias $I$ at fixed $B$'s. As shown in Fig. \ref{fig2}a, on increasing $I$, a sudden, non-hysteretic jump occurs in the $dV/dI$ curve, indicating the critical current value. The resulting critical current density $j_c$ for both samples is plotted versus $B$ in Fig. \ref{fig2}b. Interestingly, the continuous suppression of $j_c$  with increasing $B$ tails off prior to vanishing at a critical field $B_{c}^{j_c} = 12.8\,$T, $12.1\,$T and $11.9\,$T for samples J033, ITb1 and J038 respectively. Such a resilience of $j_c$ to the applied $B$ when approaching $B_{c2}(0) $ is reminiscent of the anomalous upturn of the $B_{c2}(T) $ line at low $T$. We also notice that the critical field values $B_{c}^{j_c}$ at which $j_c$ vanishes slightly differ from $B_{c2}(0)$ obtained in Fig \ref{fig1} ($B_{c2}(0)=13.3\,$T, $12.4\,$T and $12.6\,$T determined with the 50 \% criterion for samples J033, ITb1 and J038 respectively). The reasons for this disparity stem from the finite-resistance criterion used to determine $B_{c2}(T)$, which does not coincide with the termination of superconducting current at $B_{c}^{j_c}$.

\begin{figure}[h!]
	\includegraphics[width=0.8\linewidth]{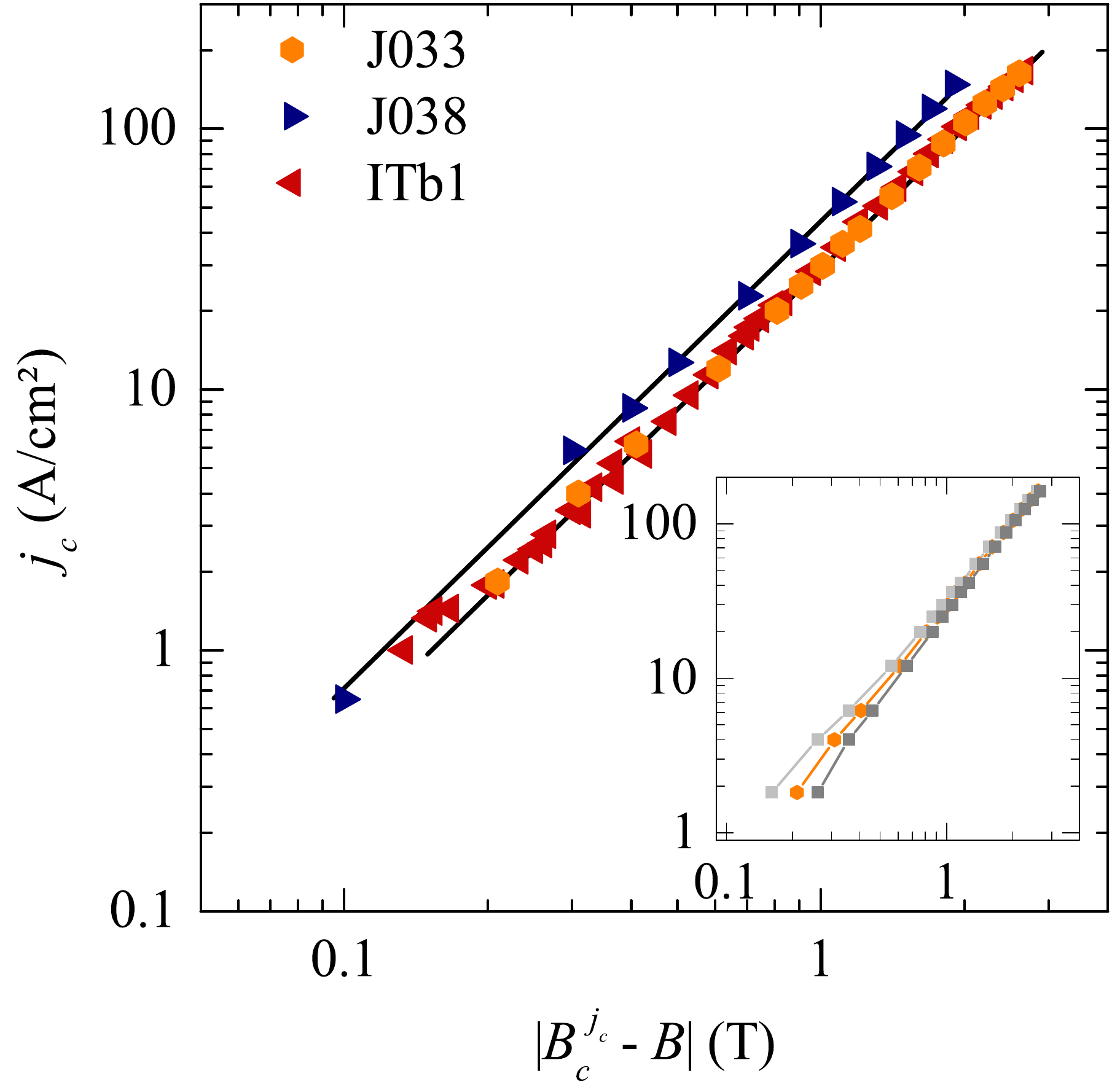}
	\caption{\textbf{Scaling of the critical current density with magnetic field.} $ j_c $ versus $|B_{c}^{j_c}-B|$. The $ B_{c}^{j_c} $ values are adjusted to obtain straight lines that are emphasized by black solid-lines. Inset: the dark grey and light grey curves are the data of sample J033 plotted with $ B_{c}^{j_c} \pm \delta $, where $ \delta = 0.05 $ T.}
	\label{fig3}
\end{figure}

The key result of this work is shown in Fig. \ref{fig3} where $j_c$ is plotted in logarithmic scale as a function of $|B_{c}^{j_c} - B|$. By adjusting the value of $B_{c}^{j_c}$ in the x-axis to $ 12.8$, $12.1$ and $11.9$ T for samples J033, ITb1 and J038 respectively, one obtains clear straight lines that unveil a scaling relation of the form: 
\begin{equation}
	j_c(B) = j_c(0)  \left|1- \frac{B}{B_{c}^{j_c}}\right|^{\upsilon},
	\label{Jc}
\end{equation}
where the fitted values for the exponent $\upsilon $ are $1.62 \pm 0.02$,  $1.67 \pm 0.02$ and $ 1.65 \pm 0.02 $ for samples J033, ITb1 and J038 respectively. The prefactor $j_c(0)$ falls in the range $(2.5 - 4) \cdot 10^3 \mathrm{A/cm^2}$ for all three samples. The inset of Fig. \ref{fig3} shows the sensitivity of the straight line to $ B_{c}^{j_c}$, where a small variation of $0.05$T yields a significant deviation from linearity. We furthermore present in the SI similar results obtained on a MoGe sample, which demonstrate the universality of this scaling relation. It is noteworthy that the values of the exponent are within 10\% of the mean-field value $ 3/2 $ of the classical temperature-driven superconducting transition in the absence of magnetic-field. Ginzburg-Landau theory predicts that the critical current would be $j_{c}^{GL} \propto \rho_s/\xi_{GL} \propto (T_c - T)^{3/2}$, with the Ginzburg-Landau superconducting coherence length $\xi_{GL} \propto (T_c - T)^{-1/2}$  and $\rho_s \propto (T_c - T)$. This striking similarity suggests  that also the scaling of  $B_{c2}(T)$ and $j_c(B)$ low temperature may be captured by mean-field theory of the bulk material.

\textbf{Interpretation of experimental results within mean-field theory}

The mean-field critical exponent of $j_c(B)$ at $T=0$ near $B_{c2}(0)$ can be extracted from the Ginzburg-Landau free energy which reads:
\begin{equation}\label{FreeEnergy}
F=\alpha|\Delta(\boldsymbol{r})|^2+\beta|\Delta(\boldsymbol{r})|^4+\gamma\Big{|}\left(-i\boldsymbol{\nabla}-\frac{2e}{\hbar{c}}\boldsymbol{A}(\boldsymbol{r})\right)\Delta(\boldsymbol{r})\Big{|}^2.
\end{equation}
Within mean-field theory, the coefficient $\alpha$  strongly depends on temperature and magnetic field~\cite{GalitskiLarkin}:
\begin{equation}\label{alpha}
\alpha=\nu\left[\ln\frac{T}{T_{c0}}+\psi\left(\frac{1}{2}+\frac{eDB}{2\pi  cT}\right)-\psi\left(\frac{1}{2}\right)\right].
\end{equation}
Here,  $D$ and $\nu$ are the electron diffusion coefficient and density-of-states respectively, and $\psi(x)$ is the digamma function. $B$ in Eq. (\ref{alpha}) is  the magnetic field penetrating the superconductor. While for type-II superconductors such as our a:InO films this magnetic field can be non-uniform at low $B$, close to $B_{c2}$ the spatial fluctuations of $B$ are negligible, and the average magnetic field is equal to the externally applied one. The other two parameters of the Ginzburg-Landau functional, $\beta$ and $\gamma\propto\nu{D}$ depend only weakly on temperature and magnetic field. Consequently, Eq. (\ref{FreeEnergy}) captures the two effects of magnetic field on superconductors: The suppression of the transition point due to pair-breaking (through the parameter $\alpha$), and the diamagnetic response captured by the vector potential $\boldsymbol{A}$. Note that the expression for $\alpha$ in Eq.~\eqref{alpha} is typical for superconductors in the presence of a pair-breaking mechanism~\cite{Galitski08}.  For example, the effect of magnetic impurities can be captured by replacing the magnetic field by a term proportional to the spin relaxation rate $\Gamma$~\cite{class1}. Correspondingly, these systems have the same $T=0$ critical exponents (i.e. $\alpha\approx |1-B/B_{c2}|$ or analogously $\alpha\approx |1-\Gamma/\Gamma_c|$). {The mean-field treatment is performed under the assumption that vortices are strongly pinned. As a result, the presence of magnetic field induced vortices can be neglected, and $j_{c}$ is proportional to the depairing current. A detailed explanation of the origin and the consequences of strong pinning is given in the next section.   }

The $T=0$ limit of Eq.~\eqref{FreeEnergy} yields the magnetic-field dependence of the order parameter and the coherence length near the quantum critical point, $|\Delta(B)|\sim|B-B_{c2}|^{1/2}$ and 
$\xi_{GL}(B)\approx \xi_0|1-B/B_{c2}|^{-1/2}$ (where $\xi_0 \approx 5$ nm in our a:InO samples, see Ref.~\onlinecite{SS1}.) 
The latter, together with the superfluid stiffness $\rho_s$, determines the mean-field value of the critical de-pairing current: $j_{c}^{GL} \propto \rho_s/\xi_{GL}$. To find the superfluid stiffness, we match  the superconducting current extracted from the free energy, $\boldsymbol{j}=-c\partial{F}/\partial\boldsymbol{A}$, with the London equation, $\boldsymbol{j} = - 4\rho_se^2 \boldsymbol{A}/\hbar^2c$. We find that  
\begin{equation}
\label{Stiffness}
\rho_s(B) = \frac{12}{\pi} \rho_{s0}  \left(1 - \frac{B}{B_{c2}(0)}\right),
\end{equation}
and the relation between the superfluid stiffness and the critical current yields 
\begin{equation}\label{eq:Jc}
j_{c}^{GL}(B) = \frac{8e}{3\sqrt{3}\pi h \rho_{ s}(B)\xi_{GL}(B)} = j_{c}^{GL}(0) 
\left(1-\frac{B}{B_{c2}}\right)^{3/2}.
\end{equation}
The corresponding critical exponent $\upsilon=3/2$ is in excellent agreement with our experimental findings. The prefactor $j_{c}^{GL}(0)$ can be estimated using the experimental data of Ref.~\onlinecite{Yazdani}, where the superfluid stiffness of $20\,$nm thick a:InO films was measured. From their experimental results at low magnetic field, we estimate the critical current to be $j_{c}^{GL}(0) \sim 10^{4} \mathrm{A/cm^2}$---larger  than our experimentally  observed value only by a factor of $\sim 4$.  This is a non-trivial observation which is in contrast to a weakly pinned vortex state where the critical current is set by the de-pinning current  $j_c^\text{depin}\ll j_c^\text{GL}$. It also provides an important hint to the origin of the anomalous critical magnetic field and, as we show below, follows naturally from our theory.

To complete the comparison between the experimental results and mean-field theory, we extract the low-temperature critical field from the condition $\alpha(B,T)=0$. 
We note that within mean field theory the resistive critical magnetic field coincides with the onset of pairing, $B_{c2}$. We find $B_{c2}(0)-B_{c2}(T)\sim{T}^2$, which is inconsistent with the linear dependence $B_{c2}(0)-B_{c2}(T)\sim{T}$ that is observed experimentally (see Fig.~2). Power-laws with exponent smaller than two are known to arise in strongly correlated superconductors~\cite{Liu1989,Dorin1990,Mikitik1992,Virshup1994,Thompson2006}. In the context of high $T_c$ cuprates, the deviation from mean-field result has been attributed~\cite{Mikitik1992} to an extended region of strong fluctuations around $B_{c2}(T)$. However, in conventional superconductors, such as the a:InO films studied here, the region of strong fluctuations is small and the Ginzburg-Landau theory is expected to describe the onset of pairing at all temperatures and magnetic fields~\cite{Mikitik1992}. Indeed, mean-field theory captures the scaling of the critical temperature with magnetic field at low $B$.  This indicates that to understand the low temperature behavior of $B_c(T)$ other beyond-mean-field effects should be considered.  In particular, Ginzburg-Landau theory does not include thermal fluctuations of the vortex-glass, which are essential in the finite-temperature transition to the normal state. As we will show below, these fluctuations are the key ingredient to understanding the linear-$T$ dependence of $B_{c2}$; however, they do not change the scaling behavior of $j_{c}(B,0)$.

\textbf{Vortex-glass fluctuations}								

In low-dimensional superconductors the pairing instability is known to differ from the onset of phase coherence. A prominent example is the Berezinskii-Kosterlitz-Thouless (BKT) transition in thin films~\cite{Berezinsky,KT}. Similar decoupling is known to occur in moderately-disordered type-II superconductors near $B_{c2}$, where the magnetic field gives rise to the formation of  a weakly pinned  vortex lattice~\cite{LO,Blatter}.  Moreover, the superconducting state becomes resistive when the force applied on the vortex lattice by the current exceeds the pinning forces. The corresponding de-pinning critical current $j_c^\text{depin}$ is then significantly lower than the pair-breaking critical current extracted from the Ginzburg-Landau theory, $j_c^{GL}$, and it is not expected to obey simple scaling behavior~\cite{Blatter,Kwok2016} close to $B_{c2}$. 

In contrast, in highly disordered superconductors such as our a:InO films, we expect a strongly pinned vortex-glass~\cite{Fisher1991,Blatter} to form. This is caused by large spatial fluctuations of the order parameter~\cite{Sacepe11}, which are predicted by the theory of ''fractal'' superconductors~\cite{FIKC2010} to arise in disordered systems such as the a:InO films in question. According to this theory,~\cite{FIKC2010} in the absence of a magnetic field, the superconducting condensation energy $E_g$ fluctuates strongly in space, $\delta E_g \sim E_g$, over distances comparable to the coherence length $\xi$. Correspondingly, core energies of vortices, which are induced by an applied magnetic field, exhibit similar fluctuations, and hence become strongly pinned. In fact, vortices pinning in such systems resembles the one found in models of columnar defects~\cite{MkrtchyanShmidt}.  Upon applying a current, vortices de-pin only when the superconducting order parameter is sufficiently reduced, i.e.,  within mean-field theory $j_c^\text{depin}$ scales like the Ginzburg-Landau de-pairing current. Thus,  in our system  $j_c^\text{depin} = \Upsilon j_c^{GL}$ where $\Upsilon < 1$  (for example, in the model of Ref.~\cite{MkrtchyanShmidt}, $\Upsilon \approx 1/3$), and the I-V curves are expected to follow those studied theoretically in Ref.~\onlinecite{Buchacek}. Consequently, Eq.~(\ref{eq:Jc}) still applies even in the presence of vortices. 

The above analysis implies that the main corrections to the mean-field values of the critical field and current in highly disordered superconductors are due to renormalization of the superfluid stiffness caused by fluctuations of the vortex glass~\cite{Fisher1991}. To estimate the modified $\rho_s$, we focus exclusively on phase-fluctuations of the order parameter, i.e, $\Delta(\boldsymbol{r})=|\Delta_0|e^{i\Phi(\boldsymbol{r})}$.
Inserting this into the free energy, Eq.~\eqref{FreeEnergy}, yields $F[\Phi]=\rho_s\left[\boldsymbol{\nabla}\Phi(\boldsymbol{r})-\frac{2e}{\hbar{c}}\boldsymbol{A}(\boldsymbol{r})\right]^2/2.$ It is convenient to further separate $\Phi(\boldsymbol{r})$ into smooth phase fluctuations (the superfluid mode) $\varphi(\boldsymbol{r})$  with $\boldsymbol{\nabla}\times \boldsymbol{\nabla}\varphi=0$  and fluctuations of the vortex-glass $\psi(\boldsymbol{r})$, with $\Phi(\boldsymbol{r})=\varphi(\boldsymbol{r})+\psi(\boldsymbol{r})$. We determine the renormalized superfluid stiffness via the static current-current correlation function $\langle{J}_{\text{sc}}^i(\boldsymbol{r},\omega_n=0)\rangle=\int d\boldsymbol{r}\langle{J}_{\text{sc}}^i(\boldsymbol{r},0)J_{\text{sc}}^j(0,0)\rangle A^j(0,0)$, where as before $\boldsymbol{J}_{\text{sc}}(\boldsymbol{r})=-c\partial{F}/\partial\boldsymbol{A(\boldsymbol{r})}$. Since $\rho_s$ is determined by the long-wavelength properties, it is sufficient to focus on  length-scales larger than the (typical) inter-vortex spacing $a_0$. Moreover, within such a coarse-grained description, the coupling between superfluid and vortex fluctuations is local. 

We note that a charge encircling a vortex  acquires phases from both the external and the vortex field, $\oint[\vect{\nabla}\psi(\vect{r})-2e/\hbar{c}\vect{A}(\vect{r})]\cdot{d}\vect{\ell}$.  If the vortices were uniformly spaced (a vortex lattice), the two contributions to the phase would cancel at a length scale $a_0$. In the coarse grained description and in the symmetric gauge for the  vector potential, this amounts to  $\boldsymbol{\nabla}\psi(\boldsymbol{r})=\vect{A}(\boldsymbol{r})$. In a vortex-glass this cancellation is not exact. Still, near $B_{c2}$ and at the length scales of $a_0 = \sqrt{\Phi_0/B} \approx\xi_0 \sqrt{2\pi} $, the density of vortices is nearly uniform. We introduce the field $\vect{R}(\vect{r})=(R_x(\vect{r}),R_y(\vect{r}))$ that describes the deviation of the vortex positions from uniformity. $\vect{R}$ thus encodes the particular realization of the vortex-glass, and, in the absence of forces, its spatial average vanishes, $V^{-1}\int{d}\vect{r}\vect{R}(\vect{r})=0$.  It is thus appropriate to expand 
\begin{align}
\boldsymbol{\nabla}\psi\left(\boldsymbol{r}-\vect{R}(\vect{r})\right)-\frac{2e}{\hbar{c}}\vect{A}(\boldsymbol{r})&\approx-\frac{2e}{\hbar{c}}\left(\vect{R}(\vect{r})\cdot\vect{\nabla}\right)\vect{A}\\\nonumber
&=\frac{\vect{u}(\vect{r})\times\hat{z}}{2a_0}.
\end{align}
where $\boldsymbol{u}(\boldsymbol{r}) = 2\pi \boldsymbol{R}(\boldsymbol{r})/a_0$ is the dimensionless displacement field.  It follows that the leading coupling terms in the free energy between  $\varphi$ and $\vect{u}$ are
\begin{align}\label{Coupling}
\delta{F}=\rho_sa_0^{-1}\hat{z}\cdot\left[\vect{\nabla}\varphi(\vect{r})\times\vect{u}(\vect{r})\right]-C\rho_s\left(\vect{\nabla}\varphi(\vect{r})\right)^2
\vect{u}^2(\vect{r}),
\end{align}
where $C$ is a material specific coefficient of order unity. Note that here we assume isotropy in the $x-y$ plane. 

The first term in Eq.~\eqref{Coupling} corresponds to the lowest order expansion with respect to gradients in the Ginzburg-Landau free energy~\eqref{FreeEnergy}. It gives rise to a temperature and magnetic-field independent correction to the superfluid stiffness that depends on the realization of the vortex-glass. This reduction enters the measurable quantity $\rho_{s0}$ which we treat as a phenomenological parameter.  As we show below, higher-order gradients (like the second term in Eq.~\eqref{Coupling}) become important at non-zero temperature. The leading contribution of such terms to $\rho_s$ is given by
 \begin{align}\label{DeltaRho}
\delta\rho_s^{x,y}&= -\frac{C\rho_s}{a_0^3}T\sum_n\int{d\vect{r}}\langle{\vect{u}(\vect{r},\omega_n)\cdot\vect{u}(0,-\omega_n)\rangle}.
\end{align} 
It remains to evaluate the $\vect{u}-\vect{u}$ correlation function. In the strong pinning regime, a restoring force  acts to keep the vortex structure near its local energy minimum. Consequently, it is sufficient to reduce the equation of motion to its local form  for
 $\boldsymbol{u}(\vect{r},t)$, which describes individual vortices~\cite{Blatter}, and in addition neglect spatial gradients of the $\boldsymbol{u}(\vect{r}) $ field 
\begin{equation}
\eta\partial_t\boldsymbol{u}(\vect{r},t) + \kappa (\boldsymbol{u}(\vect{r},t)-\boldsymbol{u}_0(\vect{r})) = \boldsymbol{f}(t) \equiv \frac{ha_0}{2e}
\boldsymbol{j}\times \hat{z},
\label{eqmotion}
\end{equation}
where  $\boldsymbol{u}_0(\vect{r})$ is the static displacement at zero current. The r.h.s. of Eq.(\ref{eqmotion}) is the  Lorentz force  acting on a segment of a vortex of length $a_0$ in presence of a supercurrent $\boldsymbol{j}$. We emphasize the absence of gradients of $\vect{u}$ in the equation of motion---this manifests the locality of the vortex dynamics. Effectively, the vortex fluctuations in the glass state at low temperature resemble a (damped)  optical phonon mode. The parameter 
$$\eta= (h/2e)^2 \sigma_n/2\pi  = \frac{\pi \hbar^2}{2 e^2} \sigma_n $$
 reflects the presence of normal electrons in the vortex cores, whose resistance $\sigma_n^{-1}$ gives friction to the vortex motion (note that we defined friction coefficient $\eta$ w.r.t. dynamics of dimensionless coordinate
$\mathbf{u} = 2\pi \mathbf{R}/a_0$).
 For strongly disordered materials $\sigma_n a_0 \sim e^2/\hbar$ and thus $\eta \sim \hbar/a_0$.  
This overdamped character of  the vortex motion leads us to neglect the inertial term $\propto \partial^2_t\boldsymbol{u}$ with  respect to friction. Likewise, we neglect the contribution to the Lorentz force due to vortex velocity since it does not affect the relevant  current-current correlation function. The parameter $\kappa$ can be determined similar to  the penetration length in pinned vortex systems, also known as  the Campbell length~\cite{Cmpb1,Cmpb11,Cmpb2} (for a recent review see Ref.~\cite{Cmpb3}).  According to  Eq.(\ref{eqmotion}), the  shift of $\boldsymbol{u} $ due to current $\boldsymbol{j}$ is equal to $\boldsymbol{u}(\vect{r})-\boldsymbol{u}_0(\vect{r}) =  {ha_0}(\boldsymbol{j}\times \boldsymbol{n} )/{2e\kappa}$.
The corresponding shift of the vortex magnetic flux can be expressed through $\delta \boldsymbol{A} = -\hbar\Phi_0\vect{j}/4e\kappa{a}_0$.
Using the London relation, we get  $ \kappa = \pi\rho_s$. From Eq.~\eqref{eqmotion} we obtain the Matsubara Green's function of $\vect{u}$ 
\begin{equation}\label{GF}
G(\vect{r},\omega_n\neq0)=a_0^2\delta(\vect{r})[\eta|\omega_n|+\kappa]^{-1}.
\end{equation}

We thus arrive at the reduction of the superfluid stiffness due to thermal fluctuations of the vortex-glass. Combining Eqs.~\eqref{DeltaRho} and~\eqref{GF} yields the following correction to the superfluid stiffness
$\delta\rho_s(T,B) = \rho_s(T,B) - \rho_s(0,B)$\, :
\begin{align}
\label{DeltaStiffness}
&
\delta\rho_s(T,B) = 
-\left[T\sum_n[\eta|\omega_n|+\kappa]^{-1}
\right.\\\nonumber
&\left.
-\int\frac{d\omega}{2\pi}[\eta|\omega|+\kappa]^{-1}\right]= -C \frac{\hbar\sigma_n}{e^2}\frac{T^2}{3\pi\rho_s(B) a_0}.
\end{align}
The last equality holds in the low-temperature limit  $T/\hbar \ll \kappa/\eta$;  in the opposite limit one recovers the classical result $\langle  \boldsymbol{u}^2   \rangle = T/\kappa$.  
The smallness of the low temperature result $\delta\rho_s(T,B)\propto T^2$ is compensated by the small $\rho_s(B) \propto 1 -{B}/{B_{c2}(0)}$  in the denominator, see Eq.(\ref{Stiffness}). This new result is unique for disordered superconductors, in which the fluctuations at $T>0$ are controlled by the dissipation in the gapless vortex cores, and it provides the key to understanding the low-$T$ linear upturn of $B_{c2}$. 

\textbf{Thermal fluctuation corrections to the critical current}

\begin{figure}[h!]
	\includegraphics[width=1\linewidth]{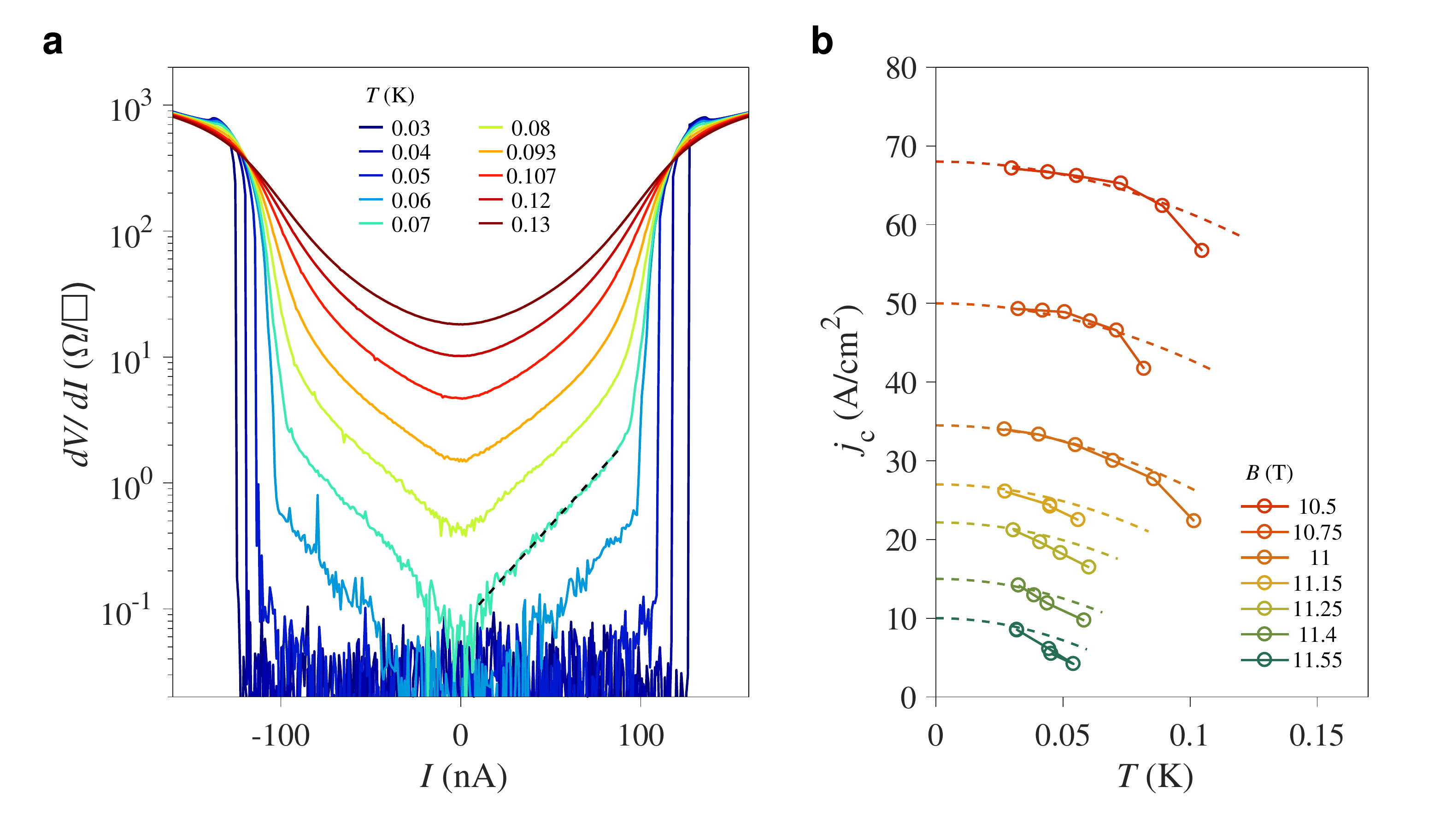}
	\caption{\textbf{Vortex de-pinning and thermal creep.} The differential resistance $ dV/dI $ versus $ I $ of sample IT1b measured at $B=11.25\,$T at different temperatures is plotted in panel \textbf{a}.  The black dashed line indicates the exponential increase of the differential resistance with the current, which is caused by vortex creep.  The critical current density $j_c $ as a function of $ T $  and $B$ is shown in panel \textbf{b}. Each value of $ j_c (B)$ was extracted from the differential resistance curves, similar to the one shown in (a), by finding the  threshold to the high resistance state. Dashed lines are fits using Eq. (\ref{TcorrectionToIc}) with $B_{c2}(0) = 12\,$T and adjusting  $j_c (T = 0, B)$ for each curve using a single prefactor for all $B$'s.}
	\label{fig4}
\end{figure}

To substantiate the correction to the superfluid stiffness given in Eq.~(\ref{DeltaStiffness}), we conducted additional measurements of the temperature dependence of the critical current in the vicinity of $B_{c2}(0)$.  The differential resistance $dV/dI$ as a function of current measured on sample ITb1 at various temperatures and fixed $B=11.25\,$T is shown in Figure \ref{fig4}a. A clear jump in the resistance, similar to that in Fig.~\ref{fig1}a,  develops at ultra-low temperatures and indicates the position of $j_c$. At higher temperatures, the critical current decreases and a non-zero resistance is found already before the jump. This resistance at sub-critical current rises above the noise level for $T>0.05\,$K,  and exhibits a clear exponential increase with the current, which is highlighted by the black dashed-line in this semi-log plot. Such resistance curves are expected to be observed when the vortex glass is strongly pinned: The resistance at low current is a typical signature of vortex creep, where the Lorentz force induced by the current reduces the barrier. Above $j_c$, the corresponding current-voltage characteristics show an excess current~\cite{Strnad64,Xiao02} (see SI), which is a signature that thermal creep persists there, in agreement with recent strong pinning theory~\cite{Thomann12,Buchacek}. 
At temperatures above $0.07\,$K strong thermal fluctuations causes the sharp jump to be replaced by a smooth evolution between the low and high resistive states~\cite{Xiao02}. Moreover, the resistance jump that is seen here only at very low $T$ points to a collective de-pinning of the vortex-glass. Notice that Joule overheating is in play here but mainly in the resistive state above the critical current (see detailed analysis in SI).

As we showed before, the (zero-temperature) $B$-dependence of the critical current scales like $j_c^{GL}$ near $B_{c2}(0)$, indicating that  $j_c^{\text{depin}}\propto j_c^{GL}$. Correspondingly,  we expect the $T$-dependence of the critical current to be determined by the thermal corrections to the superfluid stiffness
\begin{equation}\label{TcorrectionToIc}
\delta j_c^{GL}(T,B) \propto \frac{\delta\rho_s(T,B) }{ \xi_{GL}}\propto \frac{T^2}{\sqrt{B_{c2}(0)-B}}
\end{equation}
To test this predicted scaling of $j_c$ as a function of temperature, we measured additional resistance curves, similar to those shown in Fig.~\ref{fig4}a, at different magnetic fields (see SI). The critical current values near $B_{c2}(0)$ extracted from the jump in the resistance are plotted as a function of temperature in Fig.~\ref{fig4}b.  The dashed lines are fits of the $j_c(T,B)$ data to Eq. (\ref{TcorrectionToIc}), which were performed by setting $B_{c2}(0)=12.1\,$T (deduced from Fig.~\ref{fig3}), by adjusting the $T=0$ value $j_c(0,B)$ for each $B$ and finding one global  pre-factor. We note that the resulting $j_c(0,B)$ values scale as $(B_{c2}-B)^{3/2}$. The fit reproduces remarkably well the $T$-dependence of the data for the $B=10.5, 1.75$ and $11\,$T, confirming the $T^2$ correction to the critical current as well as its $B$-dependence. Deviations from the fit occur for the highest $T$ data points as well as in the immediate vicinity of $B_{c2}(0)$. This is not surprising since our theoretical derivation of the correction to the superfluid stiffness given in Eq.~(\ref{DeltaStiffness}) is valid so long as the fluctuations are small $\delta\rho_s(B,T)\ll\rho_s(B)$.  The thermal fluctuations, however,  become strong at lower $T$ as the magnetic field is increased. 

The excellent agreement between the vortex-glass fluctuation correction and the data has important implications: Together with the observation of the Ginzburg-Landau scaling of the critical current shown in Fig.~\ref{fig3}, this analysis confirms that the collective vortex depinning, which causes the jump in the resistance, is indeed proportional to the de-pairing critical current $j_c^{GL}$. Moreover, this result validates our prediction for the renormalization of the superfluid stiffness by thermal fluctuations of the vortex glass, and suggests that this should affect other observables such as magnetotransport near the quantum phase transition at $B_{c2}$.  In the following we show these fluctuations can account for the linear upturn of $B_{c2}(T)$ at low temperature.

\textbf{Theory for the low-temperature anomaly}

The superconducting transition in the bulk limit can be estimated from the condition $\rho_s(T,B) \approx0$, or equivalently when 
$\delta\rho_s(T_c,B) = \epsilon \rho_s(0,B)$, with $\epsilon$  being a number of the order unity (similar to the Lindemann criterium for melting of solids).
Under this condition and using Eqs.~(\ref{Stiffness}) and (\ref{DeltaStiffness}) we find 
\begin{equation}
1-\frac{B_{c2}(T)}{B_{c2}(0)} = C_1\frac{\pi T}{24\rho_{s0}a_0\epsilon^{1/2}},
\label{Tclocal}
\end{equation}
where $C_1^2=4C\hbar\sigma_na_0/3\pi e^2$  is of order unity for our a:InO films. In films, the transition temperature is set instead by the BKT condition:
\begin{equation}
 \rho_s(B,T_\text{BKT}) = \frac{\chi}{d}T_\text{BKT} (B) 
\label{BKT}
\end{equation} 
where $d$ is the film thickness and $\chi^{-1}$ was numerically found~\cite{Schmidt} to be between 1.5 to 2.2 (this holds for \textit{any value of} $d$~\cite{Schmidt,Ambegaokar,Williams,Schultka}).  Moreover, kinetic inductance measurements of thin a:InO films have observed  a universal jump in the two-dimensional superfluid density per square even in the presence of a magnetic field, indicating that the BKT transition persists near the quantum critical point~\cite{Yazdani}. We note that Eq.~(\ref{BKT}) in the limit $d\rightarrow\infty$  reproduces the condition for the bulk transition  [$\rho_s(T,B) \approx0$], and thus this equation describes the scaling  of $B_c(T)$ for any $d$.  Thus, combining the BKT condition with the renormalized $\rho_s$  given by Eq.~\eqref{DeltaStiffness}  yields the scaling of the transition temperature with magnetic field as a function of thickness
\begin{equation}
1-\frac{B_{c2}(T)}{B_{c2}(0)} = \left[1+\sqrt{1+C_1^2\frac{d^2}{\epsilon a_0^2\chi^2}}\right]\frac{\pi\chi{T}}{24\rho_{s0}d}.
\label{final}
\end{equation}
We thus obtain a linear temperature dependence of $B_{c2}(T)$ at low $T$, which  provides a theoretical description of our experimental data shown in Fig.~\ref{fig1}b. In addition, this result is in agreement with numerous experiments in films~\cite{Tenhover81,upturn1,upturn2,Graybeal84,upturn3,upturn4,GaChina,SS1} and bulk~\cite{Bustarret04,Ren13} disordered superconductors.

We emphasize that the linear-$T$ dependence of $B_{c2}(T)$ does not depend on the sample dimensionality and, in particular, remains valid for bulk superconductors. For films, $|\delta\rho_s(B,T)/\rho_{s}(B)|$ grows with the film thickness $d$, however it remains much below unity  so long as  $d\ll a_0$ (As is the case for the experiments reported in Ref.~\cite{GaChina}). In this thin limit our theory predicts that the slope $-dB_{c2}(T)/dT$ grows linearly with $(\rho_{s0}d)^{-1}$. For thicker films, however, $|\delta\rho_s(B,T)/\rho_{s}(B)|$ increases and higher order corrections to $\rho_{s}(B)$ can become important. Still, $dB_{c2}(T)/dT|_{T\rightarrow0}$ maintains the structure $(\rho_{s0}a_0)^{-1}(g_0+g_1{a_0}/d)$ with numerical coefficients $g_{0,1}$ that (may) deviate from those given in Eq.~\eqref{final}. To conclude, while our analysis provides an exact expression for the slope only in the thin-film limit, it predicts a distinct $d$-dependence that holds for any thickness.

\begin{figure}[h!]
	\includegraphics[width=0.9\linewidth]{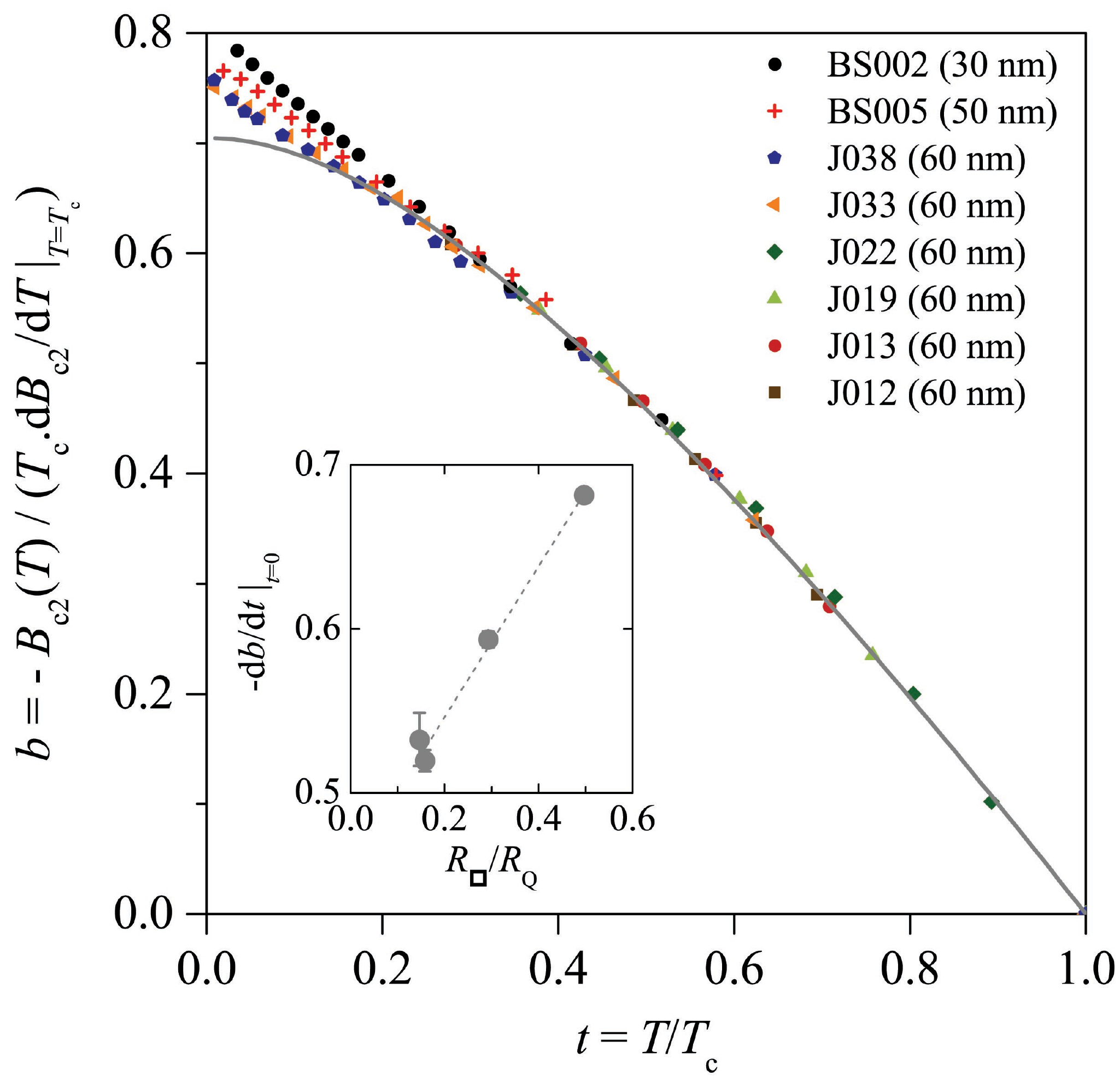}
	\caption{\textbf{Disorder dependence of the low-$T$ anomaly.} $b=B_{c2}(T)/(-T_c.dB_{c2}/dT|_{T=T_c})$ versus reduced temperature $t=T/T_c $ for 8 samples of different thicknesses. The solid gray line is a fit using the mean field theory for dirty superconductors emphasizing the deviations at low $T$. Inset: Slope $-db/dt$ at zero temperature versus $R_{\square}/R_Q$.}
	\label{fig1bis}
\end{figure}

Finally, we provide a quantitative comparison between the theoretical prediction and experimental data from eight samples of various thicknesses and resistances.  Experimentally, the physical parameters controlling the low-temperature linear deviation are difficult to assess directly from the $B_{c2}(T)$ curves. The reduction of $T_c$ and increase of $B_{c2}(0)$ upon increasing sheet resistance leads to an apparent increase of the slope $ dB_{c2}/dT|_{T\rightarrow0}$ in the $B-T$ plane  (see fig. 2b in ref.~\onlinecite{SS1}). This makes a direct comparative analysis between different samples impossible. The non-universal dependence of $T_c$ and $B_{c2}(0)$ on disorder can be eliminated by considering the quantity $b(t)=B_{c2}(T)/(-T_c\, dB_{c2}/dT|_{T=T_c})$ versus the reduced temperature $t=T/T_c $. Figure \ref{fig1bis} displays $b(t)$ for the eight different  samples  together with the theoretical mean-field curve, in solid line, found by setting $\alpha $ to zero~\cite{class1,class2}.  We see that, for $t\gtrsim 0.2-0.3$, all high-temperature data collapse on the theoretical curve. At smaller $t$, the low-temperature anomaly of $B_{c2}$ develops as a linear deviation from the mean-field curve. Our theory explains the anomalous slope in this regime and, as we show below,  captures the dependence on the sample parameters. 

To compare the plot in Fig.~\ref{fig1bis} with our theory, we extract $b(t)$ from Eq.~\eqref{final} and find:
\begin{align}\label{limits}
-\frac{db}{dt} = \begin{cases} KR_{\square}/R_Q & d\ll a_0 \\
\tilde{g}_0 \sqrt{\frac{1}{\sigma_nR_Q a_0}} + \tilde{K}\frac{R_{\square}}{R_Q} & d\gg a_0
 \end{cases},
\end{align}
where $R_{Q}=h/4e^2$ is the quantum of resistance for electron pairs (for details see the Supplementary Information). Neglecting higher order corrections to $\rho_s$ beyond Eq.~\eqref{final}, and the suppression of $\rho_{s0}/T_{c0}$ in strongly disordered superconducting films yields $K=0.4\pi \approx 1$,  $\tilde{K} \approx 0.1$, and $\tilde{g}_0 \approx 0.05\sqrt{C}$.   The slopes $db(t)/dt|_{t=0}$ of four samples: $30\,$nm, $50\,$nm, and two $60\,$nm thick films  (samples BS002, BS005, J033 and J038 respectively) are shown in the inset of Figure~\ref{fig1bis}. We clearly see that the slope of the linear-$T$ anomaly increases with sheet resistance as predicted by Eq.~\eqref{limits}, demonstrating the consistency of our theoretical description. Furthermore, the linear dependence of  $db(t)/dt|_{t=0}$ on the sheet resistance does not extrapolate to zero in the bulk limit $(d\rightarrow\infty)$, confirming again our theoretical result Eq.~\eqref{limits}. However, fitting the data to Eq.~\eqref{limits} gives $-db/dt= 0.4R_{\square}/R_{Q}+0.44$,  i.e., $\tilde{g}_0\approx0.45$  and $\tilde{K}\approx0.4$.  These values exceed the ones found via our simplified approximation by a factor of about 4-10. Partially, it might be due to the overestimation of  the ratio $\rho_{s0}/T_{c0}$, see Supplementary Information.
 Obtaining the correct numerical coefficients presumably requires including corrections neglected above, which is beyond the scope of this work.

In conclusion, we have conducted a systematic study of the critical current near $B_{c2}(T=0)$ in disordered a:InO films. These films are prototypical representatives of a variety of  superconducting film in which the critical field shows an anomalous linear upturn as $T\rightarrow0$. We uncovered a power-law dependence of $j_{c}$ on magnetic field and extract the corresponding critical exponent. This previously unknown scaling property elucidates the origin of the 'critical field anomaly'---the strong deviation of $B_{c2}$ from the mean-field result at the lowest temperatures. We have showed theoretically that  the behavior of \textit{both} $B_{c2}(T)$ and $j_{c}(B)$ can be attributed to the specific properties  of  the vortex-glass, which is characteristic state of disordered films in the presence of magnetic field. 
In particular, the anomalous dependence of  $B_{c2}$ on  $T$ is induced  by a soft overdamped (bosonic) fluctuation mode of the vortex-glass. Although this mechanism is quite generally applicable for any  disordered superconductor, both 2D and 3D, the magnitude of the effect is appreciable for superconductors with low superfluid density, $\rho_s \xi_0 \leq T_{c0}$, where phase fluctuations are strong~\cite{EK1995}.

Our analysis provides sharp predictions for $B_{c2}(T)$ and $j_c(B)$ which allows a clear distinction between the physical mechanisms in play. It would be very interesting to look for similar scaling in other superconducting films (in particular amorphous/non-amorphous) where a similar $B_{c2}$ anomaly has been observed~\cite{Tenhover81,upturn1,upturn2,Graybeal84,upturn3,upturn4,Bustarret04,Ren13,GaChina}.  
A detailed measurement of the thickness and field dependence of the critical current could shed light on difference and similarities of the vortex-glass phases in conventional and also unconventional superconductors.

\section*{Methods}

\textbf{Sample fabrication and measurement setup.}
Disordered a:InO films were prepared by e-gun evaporation of 99.99 \% In$ _2 $O$ _3 $ pellets on a Si/SiO$ _2 $ substrate in a high-vacuum chamber with a controlled O$ _2 $ partial pressure. Films were patterned into Hall bar geometry ($100 \,\mu$m wide for all samples except ITb1 which is $20\,\mu$m wide) by optical lithography, enabling four-terminal transport measurements using standard low-frequency lock-in amplifier and DC techniques.  Measurements were performed in dilution refrigerators equipped with superconducting solenoid. Multi-stage filters, including feed-through $\pi$-filters at room temperature, highly dissipative shielded stainless-steal twisted pairs down to the mixing chamber, copper-powder filters at the mixing chamber stage, and $47\,$nF capacitors to ground on the sample holder, were installed on each dc lines of the fridge.  This careful filtering of low and high frequency noises was crucial to measure well defined critical currents as low as $\sim 50\,$nA (see Fig. \ref{fig4}), which would have been otherwise disguised by spurious noise. Furthermore, a calibrated RuO$_2$ thermometer was installed directly on the sample holder to precisely monitor the sample temperature. This accurate thermometry eliminates small temperature gradients below $0.1\,$K.

\vspace{1cm}

\textbf{Acknowledgments} We are grateful to Vadim Geshkenbein, Lev Ioffe, Thierry Klein and Mikhail Skvortsov for useful discussions. We thank Idan Tamir and Dan Shahar for providing sample ITb1. This work was supported by the Lab Alliance for Nanosciences and Energies for the Future (Lanef) and the H2020 ERC grant \textit{QUEST} No. 637815. Research of M.V.F. was partially supported by the Russian Science Foundation grant \# 14-42-00044.

\vspace{1cm}



\clearpage
\onecolumngrid
\setcounter{figure}{0}
\setcounter{section}{0}
\renewcommand{\thefigure}{S\arabic{figure}}


\begin{center}
\textbf{\large Supplementary Information of \\ Quantum meets classical phase transition: Low-temperature anomaly in disordered superconductors near $B_{c2}$} \vspace{5mm}
\end{center}

\makeatletter\renewcommand{\thefigure}{S\@arabic\c@figure}
\makeatletter\renewcommand{\thetable}{S\@arabic\c@table}

\linespread{1.5}

\section{Sample parameters}
Table~\ref{table1} provides the thickness, sheet resistance before superconducting transition, and critical temperature determined by a $50\%$ drop in the normal state resistance for all the a:InO samples studied in this work. MoGe030 is an amorphous molybdenum-germanium (MoGe) thin film that has been deposited by magnetron sputtering from Mo78Ge22 composite target fabricated KJ Lesker company.  We used Si substrate covered with a 60-nm-thick layer of SiN grown by chemical vapor deposition. Prior to the deposition of the MoGe film, a $3$-nm-thick underlayer of amorphous Ge was deposited. This underlayer is critical for the growth of the homogeneous MoGe films~\cite{Kim12}. For oxidation protection the films were covered by a 10-nm-thick layer of Ge. All layers were deposited without breaking vacuum (base value $5\times 10^{-8}$ Torr) via the same shadow mask made in thin stainless steel plate.  

\begin{table*}[h!]
	\centering
	\begin{tabular}{lccc}
		\hline \hline
		Sample & $ d $ (nm) & $R_\square^{max} \,(k\Omega)$  & $T_c \,(K)$ \\ \hline
		J012 & 60 & 0.5  & 3.5 \\
		J013 & 60 & 0.7  & 3.4  \\
		J019 & 60  & 1.3  & 3.1 \\
		J022 & 60 & 2.3  & 2.7  \\
		J033 & 60 & 1.3  & 3.2  \\
		J038 & 60 & 1.2  & 3.5  \\
		BS005 & 50 & 2.2  & 2.6  \\
		BS002 & 30 & 3.5  & 2.4  \\
		ITb1 & 30 & 2.5  &  3.0  \\
		MoGe030 & 3 & 0.7 & 4.0 \\
		\hline \hline
	\end{tabular}
	\caption{\label{table1} Sample parameters. Thickness, $ d $, was measured by atomic force microscopy; $R_\square^{max}$ is the maximum sheet resistance reached before the superconducting transition. $T_c$ is the superconducting critical temperature.}
\end{table*}

\section{Amorphous molybdenum-germanium thin film}

\begin{figure}[h!]
	\includegraphics[width=1\linewidth]{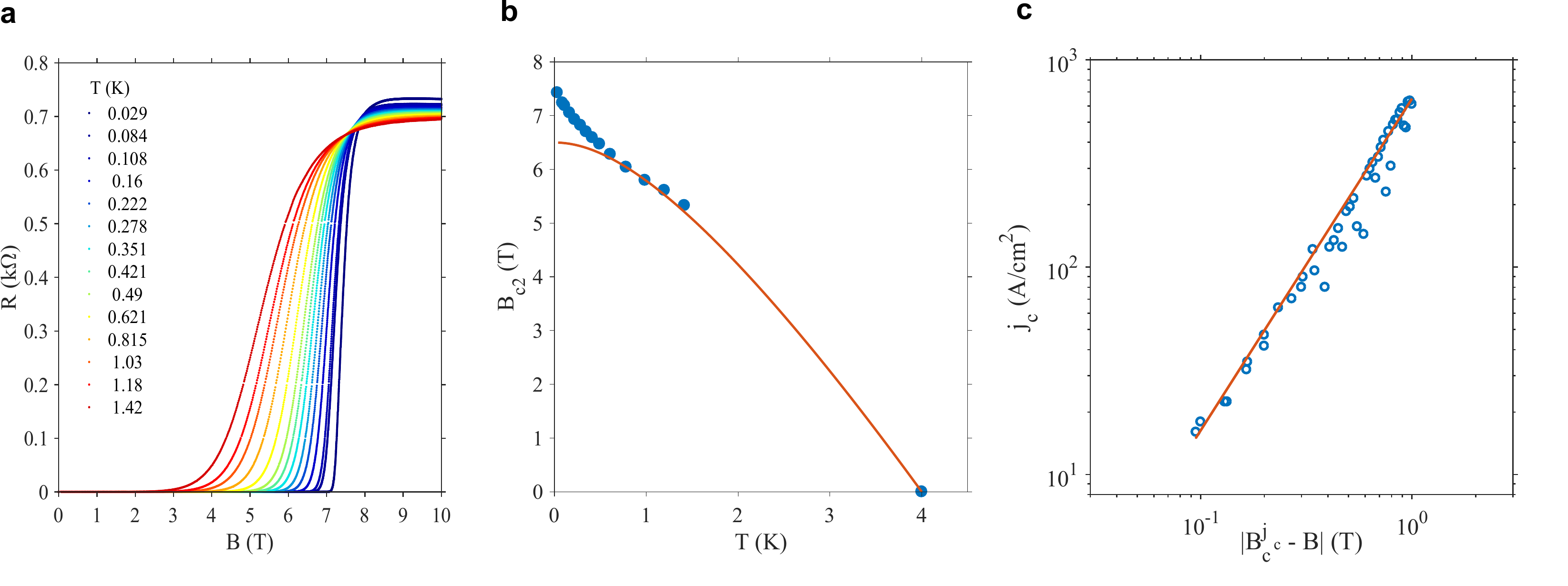}
	\caption{\textbf{MoGe thin film.} \textbf{a,} Magnetoresistance curves measured at different temperatures with an excitation current of $1$~nA. \textbf{b,} $B_{c2}$ versus $T$ extracted from \textbf{a} at $50\%$ of normal state resistance. The red line is a fit with the theory of dirty superconductors. \textbf{c,} Critical current density $j_c$ versus $B_{c}^{j_c}-B$ with $B_{c}^{j_c}=7\,$T. The red line is a guide for the eyes of slope $3/2$.}
	\label{figS-MoGe}
\end{figure}

To test the universality of our results on the anomalous dependence of the critical magnetic field on temperature, we repeated the measurements on another material, that is, amorphous molybdenum-germanium (MoGe), whose parameters are given in Table~\ref{table1}. Figure~\ref{figS-MoGe}a displays the magnetoresistance curves of this sample measured at different temperatures. Similar to the case of the a:InO samples, the transition to the normal state becomes steeper upon decreasing temperature, indicating strong vortex pinning. The temperature dependence of the critical field, $B_{c2}(T)$, extracted from these magnetoresistance data at $50\%$ of the normal state resistance is shown in Fig.~\ref{figS-MoGe}b together with a fit of the mean-field theory of dirty superconductors~\cite{class1,Tinkham}. A clear low-$T$ anomaly develops below $0.6$~K with a nearly linear increase of $B_{c2}(T)$ on lowering further the temperature.

We then performed systematic measurements of the critical current in the MoGe sample near $B_{c2}(0)$. The results are shown in Fig.~\ref{figS-MoGe}c which presents the critical current density $j_c$ as a function of $B_{c}^{j_c}-B$. For $B_{c}^{j_c} = 7\,$T, most of the data collapse on a straight line which is denoted by the red line of slope $3/2$. This collapse is in agreement with our previous experimental findings, which are reported in the main text, i.e.,  the mean-field scaling $j_c \propto (B_{c}^{j_c}-B)^{3/2}$. 
The scattering of the data, in particular below the red line, is a consequence of sample heating by the large current in the resistive state (see section IV). When the thermalization time in the superconducting state is not long enough while sweeping the current, the critical current is reduced compared to it's zero temperature limit. 

To conclude, the study of a second material consistently confirms that, in disordered thin films, the de-pinning critical current is proportional to the de-pairing current with a mean-field scaling $j_c \propto (B_{c}^{j_c}-B)^{3/2}$. Moreover, the data obtained for the MoGe film demonstrate the generality of our explanation for the origin of the low-$T$ anomaly of the critical field.

\section{Differential resistance}
\begin{figure}[h!]
	\includegraphics[width=0.8\linewidth]{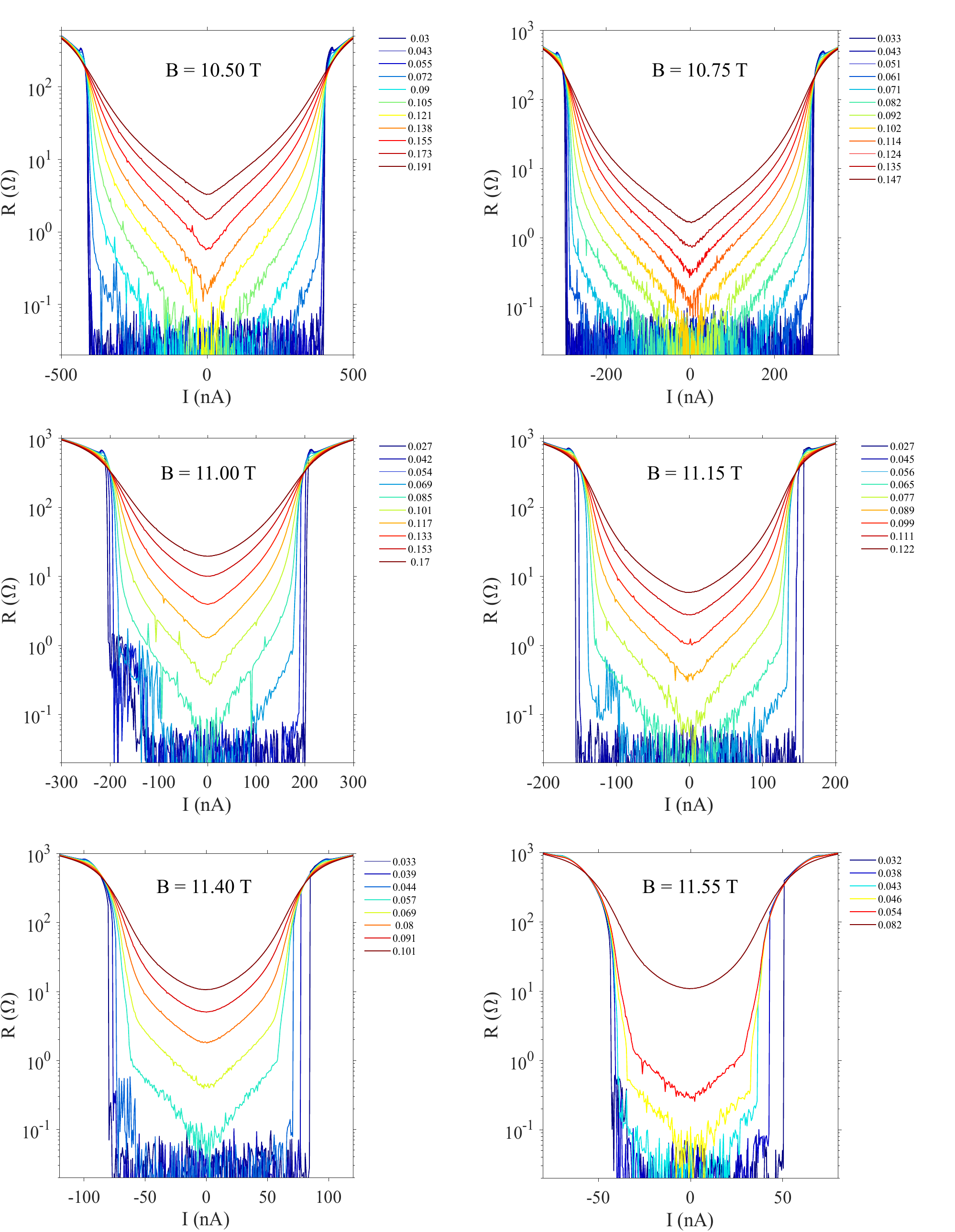}
	\caption{$dV/dI$ curves versus applied current $I$ for various temperatures and magnetic fields on sample ITb1. Current is systematically swept from negative to positive values.}
	\label{figS1}
\end{figure}

In this section we present the full set of data used to extract the temperature dependence of the critical current as shown in Fig. 4b of the main text.  Figure~\ref{figS1} displays the differential resistance $dV/dI$ curves  versus current, measured at different temperatures and magnetic fields. The applied dc current was swept from negative to positive values while measuring the differential resistance with a lockin amplifier technique and an ac current modulation of $1\,$nA. Temperature was carefully monitored with an on-chip thermometer (see Methods). For all set of curves, a resistance jump develops at low temperature for a given current threshold. This current threshold is slightly hysteretic, that is, depends on the direction of current sweep.  In this work, the critical current is identified as the current threshold on the switching to the resistive state at positive current values. Below the critical current, the differential resistance exhibits an exponential increase with the applied current. This exponential increase persists even at high temperatures when the jump is absent and replaced by a smeared crossover to the resistive state.

\begin{figure}[h!]
	\includegraphics[width=0.6\linewidth]{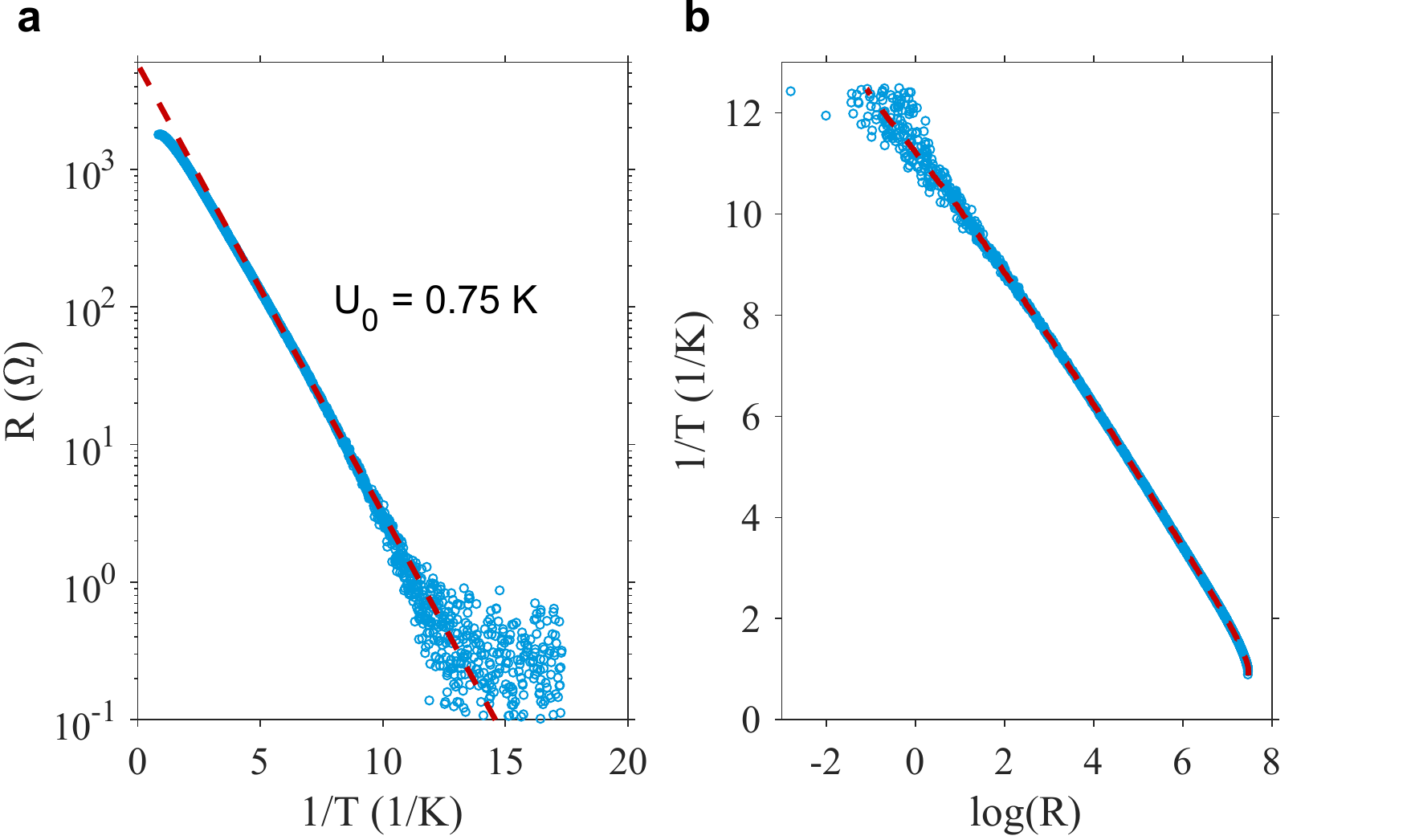}
	\caption{\textbf{Thermally assisted flux creep.} \textbf{a,} Arrhenius plot of the resistance at $B=11.25$~T. The red dashed line is an Arrhenius fit giving an activation energy of $U_0=0.75$~K and a prefactor of $5.8\,k\Omega$. \textbf{b,} Same data but with a polynomial fit that is used as the electron temperature thermometer.}
	\label{figS2}
\end{figure}

The exponential increase of the differential resistance in our films relates to vortex creep. Vortex motion is apparent in the temperature dependence of the resistance at zero bias shown in Fig.~\ref{figS2}, which exhibits an activated behavior down to our lowest temperature. This activation at zero bias results from thermally assisted flux-flow with a barrier $U(B)$ that is a function of the magnetic field. At non-zero bias, the Lorentz force applied to the vortices by the driving current reduces the activation barrier in the following way:~\cite{Anderson}
 \begin{equation} 
U(B,j) = U(B)\left(   1-\frac{j}{j_1}\right),\\
\end{equation} 
where $j_1 = \varsigma\frac{U(B)}{\xi^2dBL}$ is a current density related to the superconducting coherence length $\xi$, the thickness $d$ and the distance between two potential wells $L$ ($\varsigma$ being a constant). As a result, the differential resistance grows exponentially with \textit{both} $T$ and $j$,
\begin{equation}
R(T,j)=R_0 e^{-U(B,j)/T}.
\label{eq:Tdependence}
\end{equation}
This vortex creep picture describes very well our data in which a systematic exponential increase of the differential resistance on increasing current is invariably observed at low temperatures below the critical current.

\begin{figure}[h!]
	\includegraphics[width=0.4\linewidth]{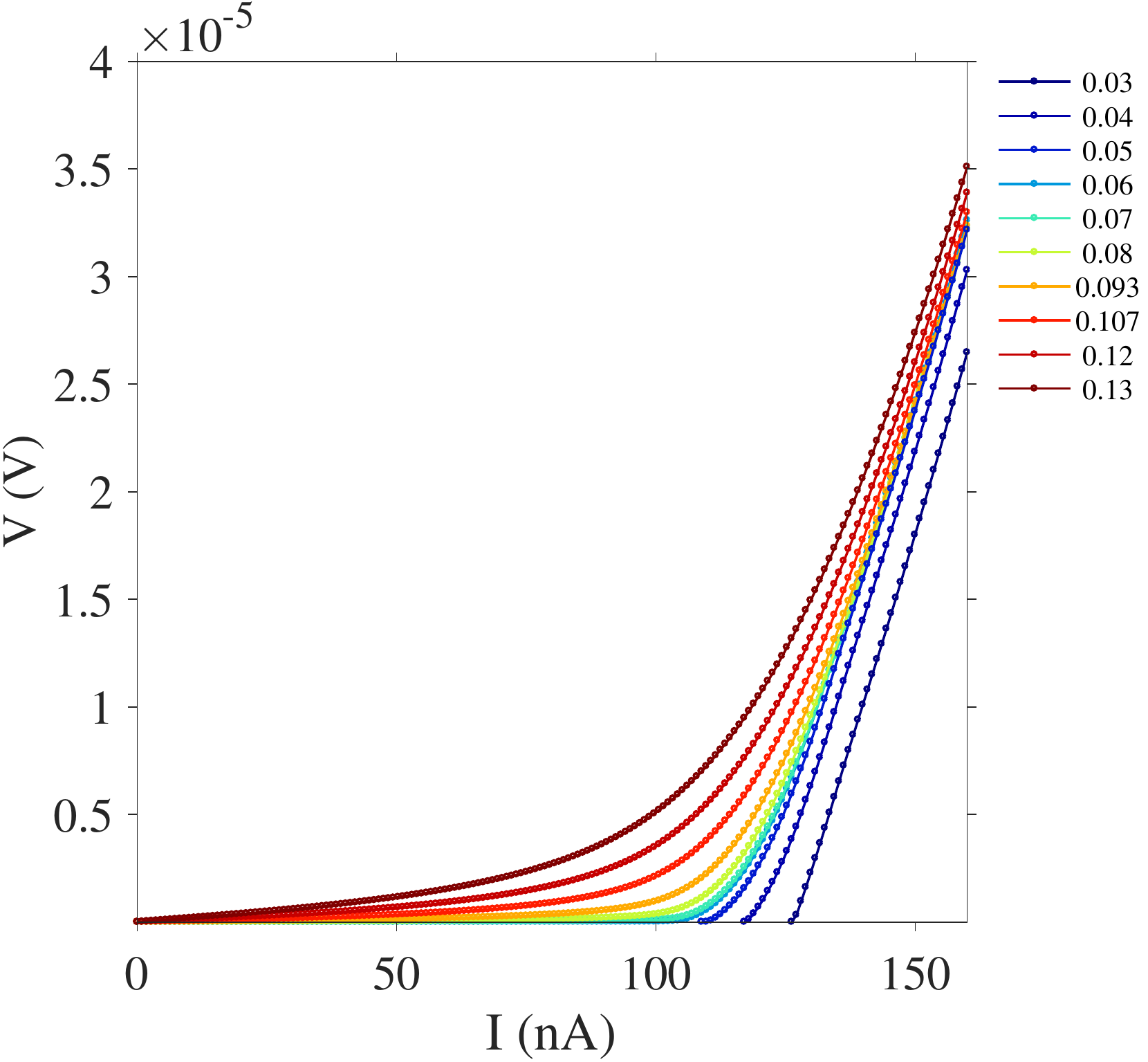}
	\caption{\textbf{Excess current.} $I$-$V$ characteristics obtained after numerical integration of the data of Fig.~4a of the main text. The resistive state extrapolates linearly at $V=0$ to the critical current.}
	\label{figS2b}
\end{figure}

To investigate the $I$-$V$ characteristics above the critical current,  we numerically integrated the differential resistance curves shown in Fig.~4a of the main text. The resulting $I$-$V$'s are shown in Fig.~\ref{figS2b}. At the lowest temperatures, a non-zero voltage sets in above the critical current and then raises linearly on increasing further the bias current. This linear $I$-$V$ curves indicate an ohmic regime of flux-flow that is shifted by an excess current offset corresponding to the critical current. Such peculiar behavior were observed earlier in various disordered superconductors~\cite{Strnad64,Kim65,Berghuis93,Xiao02} and have been recently interpreted as a signature of vortex creep persisting beyond the critical current when vortices are stronly pinned~\cite{Thomann12,Buchacek18}.

\section{Joule overheating scenario}

Joule heating is ubiquitous in non-equilibrium measurements. At low temperature, when the electron-phonon coupling is weak, dissipation due to an applied bias can potentially induce electron-overheating and subsequently, non-linear $I$-$V$ characteristics~\cite{Gurevich87}. This is particularly true with disordered conductors when the resistance is large and exhibits a diverging temperature dependence due to transition into insulating or superconducting states. In disordered insulators, such as the high field regime of the $B$-induced insulator in a:InO films, non-linearities in the $I$-$V$'s mainly stem from electron-overheating~\cite{Ovadia09,Altshuler09}. At very low $T$, when the bias voltage $V$ reaches a voltage threshold $V_{th}$, these non-linear $I$-$V$'s exhibit an hysteretic current jump of several orders of magnitudes between a high resistive state at $V<V_{th}$ and a low resistive state at $V>V_{th}$. The hysteresis and current jump have been proven to be a direct consequence of a thermal bistability of the electronic system driven by Joule overheating~\cite{Ovadia09,Altshuler09}. 

In this section we address the possibility that the resistance jumps and the accompanying non-linearities we observe near $B_{c2}(0)$ in our moderately disordered superconducting a:InO films result from a similar electron overheating and thermal bistability. For this purpose, we analyze the heat balance equation~\cite{Ovadia09,Altshuler09}:
\begin{equation}
P=R(T_{el})\times I^2=\Gamma \Omega \left(T_{el}^{\alpha} - T_{ph}^{\alpha} \right),
	\label{eq:HeatBal}
\end{equation}
where $\Gamma$ is the electron-phonon coupling constant, $\Omega$ the sample volume, $T_{ph}$ is the phonon bath temperature as measured by the on-chip thermometer, and $\alpha$ the exponent describing the electron-phonon heat transfer.  The left-hand-side of the equation describes the heat pumped into the electronic system by the applied current, while the right-hand-side corresponds to the heat dissipated to the lattice due to electron-phonon interactions. Since $R(T_{el})$ is a steep function of the temperature near $B_{c2}(T)$ (see Eq. (\ref{eq:Tdependence})), a small change in $T_{el}$ yields a large change in the resistance and hence, a significant non-linearity in the $I$-$V$. Next, we follow the analysis of Ref.~\onlinecite{Ovadia09}: Using the experimental data for $R(T_{el})$ and the $I$-$V$ curves, we find the electrons temperature as a function of the current. To extract the temperature, we assume that $R(T_{el})$ is a smooth exponential function of the temperature, as measured at low current. Finally, we check  that  the calculated $T_{el}$ satisfies the heat balance equation (Eq.~\ref{eq:HeatBal}). If the later is correct, the jump in the resistance can be attributed to heating effects, and the critical current is not a property of the superconducting state.

\subsection{Heating analysis of the data at $B=11.25\,$T}

\begin{figure}[h!]
	\includegraphics[width=0.8\linewidth]{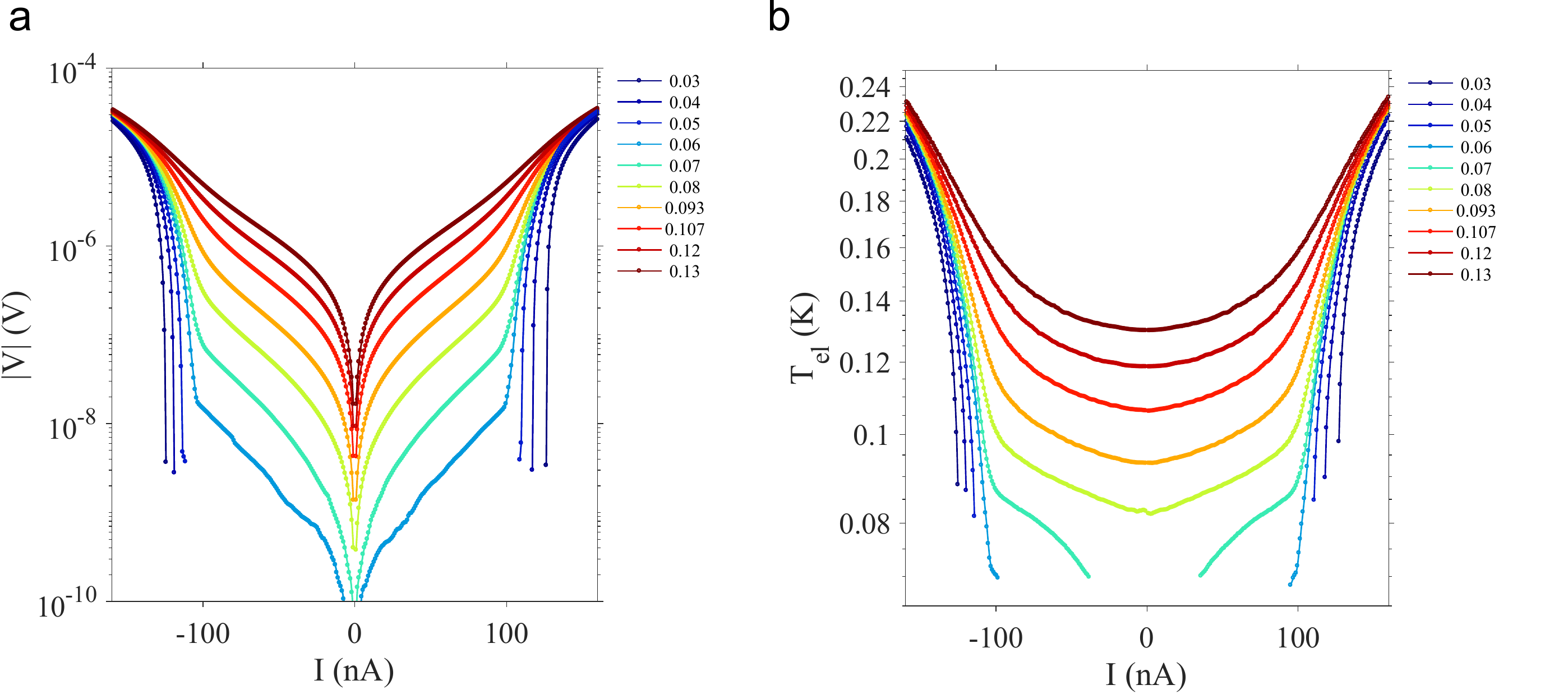}
	\caption{\textbf{Effective electronic temperature.} \textbf{a,} $|V|$ versus $I$ for different phonon temperatures. The voltage is obtained by numerical integration of the $dV/dI$ curves of Fig.~4a. \textbf{b,} Effective electronic temperature $T_{el}$ versus $I$, extracted from \textbf{a} and the measured $R(T)$ curves of Fig.~\ref{figS2}.}
	\label{figS3}
\end{figure}

We demonstrate the analysis of the heat balance equation on the data presented in Fig.~4a, which was measured at $B=11.25\,$T. We start by numerically integrating the $dV/dI$ curves to extract the voltage drop across the sample. Figure~\ref{figS3} displays the resulting $I$-$V$'s in a semilog scale for different temperatures. Notice that although no voltage jump is observed at at $T=0.06\,$K, we still find a resistance jump (see  Fig.~4a). For the lowest $T$, the integrated voltage below the current threshold is not shown since the differential resistance measurement remains below the noise level of our apparatus. A polynomial fit of the zero-bias resistance curve, $R_{I\rightarrow0}(T_{el}=T_{ph})$, shown in Fig.~\ref{figS2} is used to extract the electronic temperature as a function of the current for each $I$-$V$ curve. The resulting $T_{el}$ versus current curves are presented in Fig.~\ref{figS3}b. The effective electronic temperature raises by a factor two once the film transfer into the resistive state above the critical current.
 
\begin{figure}[h!]
	\includegraphics[width=0.8\linewidth]{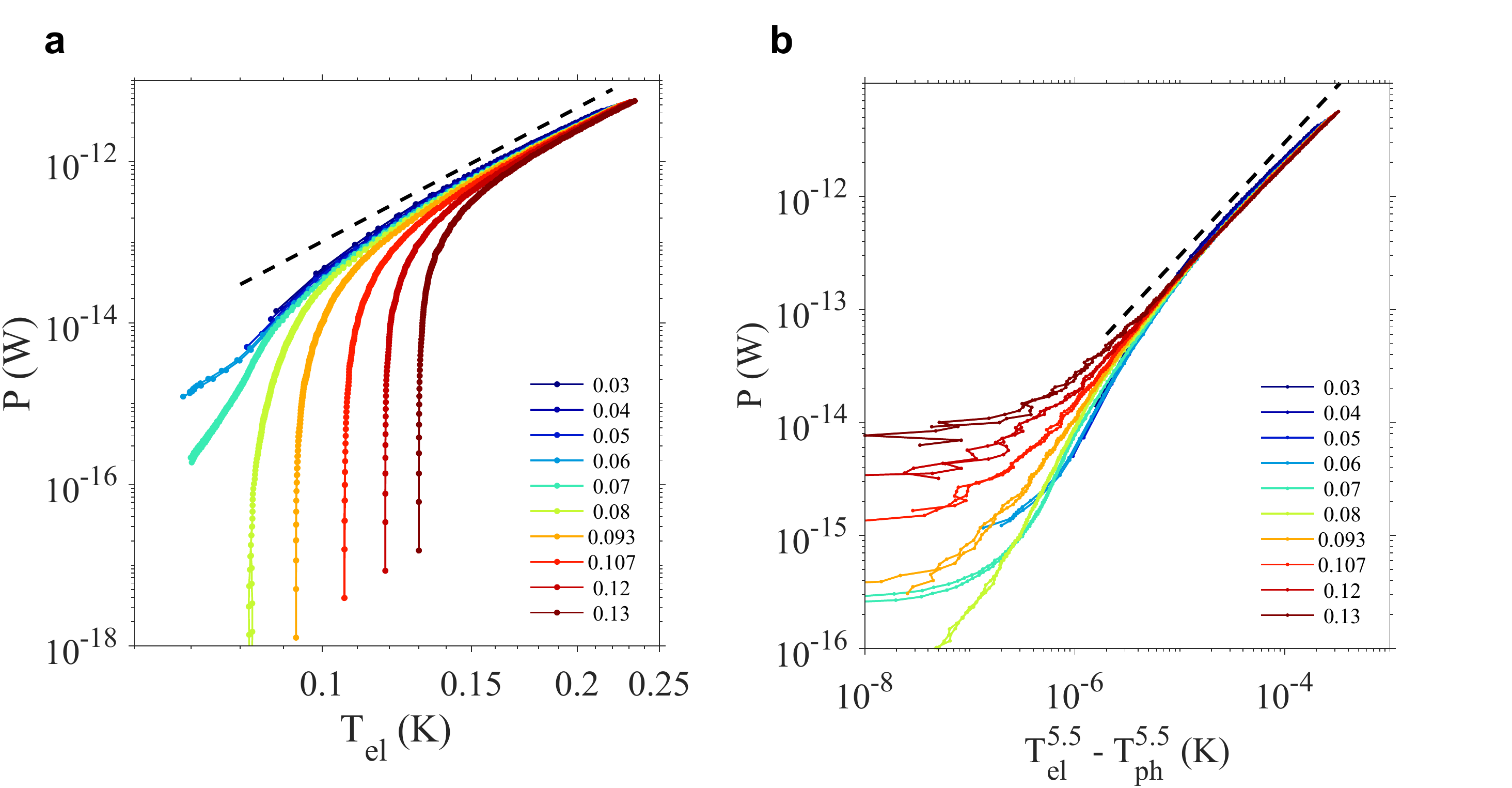}
	\caption{\textbf{Heat balance self-consistency check}. \textbf{a,} Dissipated power $P$ versus $T_{el}$ for different $T_{ph}$'s. The high $T_{el}$ and $P$ regime of these curves provides an estimate of the exponent $\alpha $ of the heat balance equation (\ref{eq:HeatBal}). \textbf{b,} $P$ versus $T_{el}^{5.5} - T_{ph}^{5.5}$.}
	\label{figS4}
\end{figure}

We turn to computing the dissipated power $P$ for each $I$-$V$, and we plot it as a function of $T_{el}$ in Fig~\ref{figS4}a. The resulting curves  collapse at high power and high $T_{el}$, i.e., in the resistive state above the critical current. Such a collapse is typically an indication that some electron overheating takes place in this regime. Furthermore, it enables us to assess the exponent $\alpha $ of the heat balance equation (\ref{eq:HeatBal}). We find $\alpha \simeq 5.5$, which is slightly different from $\alpha = 6$ found in the high field regime of the $B$-induced insulator of a:InO films near the superconductor-insulator transition~\cite{Ovadia09}.  To self-consistency check the heat balance, we plot in Fig.~\ref{figS4}b  the  power $P$ as a function of $T_{el}^{5.5} - T_{ph}^{5.5}$.  In case of a purely heating-related non-linearities, we expect a collapse of all data on a straight line of slope $1$ which confirms the heat balance equation.  The resulting plot displayed in Fig.~\ref{figS4}b displays a collapse of the data at $P>10^{-13}$~W on a straight line of slope $1$ (the dashed line is a guide for the eyes with a slope $1$) similar to that in Fig~\ref{figS4}a. Inspecting Fig.~\ref{figS3}a, we obtain that this collapse occurs at voltage and current ranges ($\geq 1\,\mu $V  and $\geq 100$~nA) that are well above the resistance jump (or the steep increase replacing the jump at high $T$). For $P<10^{-13}$~W, however, the data conspicuously does not collapse. This demonstrates the breakdown of the heating scenario and/or the assumption of intrinsic linearity of the system in the low-$P$ regime of the data. We note that small variation of $\alpha $ in this plot does not enable to recover a collapse of the low-power data. 

\subsection{Vortex creep below the critical current versus heating scenario}

\begin{figure}[h!]
	\includegraphics[width=1\linewidth]{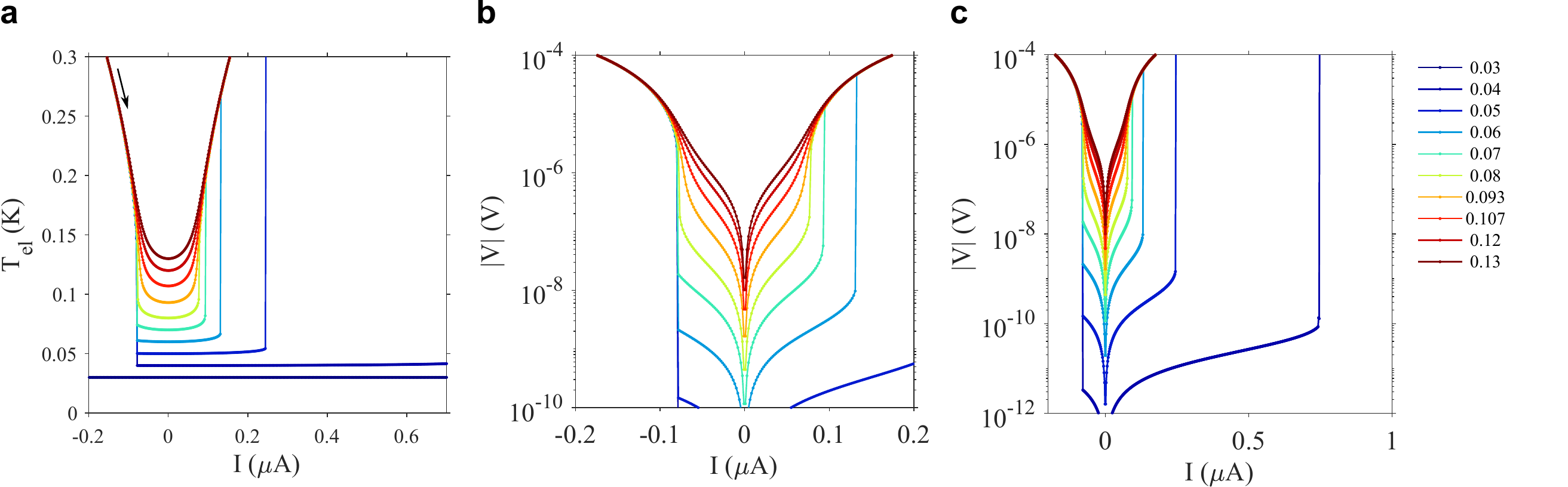}
	\caption{\textbf{Numerically computed $I$-$V$'s}.\textbf{a,} $T_{el}$ versus $I$ computed from the heat balance equation (\ref{eq:HeatBal}) and the thermally assisted flux flow equation (\ref{eq:Tdependence}) that describe the low-$T$ dependence of the zero-bias $R(T)$. The black arrow indicates the direction of sweep that defines the polarities of retrapping and switching currents. \textbf{b,} resulting $I$-$V$'s computed with $V =  R(T_{el}) \times I$. \textbf{c,} same as in \textbf{b} but in a larger scale to emphasize the divergence of the switching current at low temperature.}
	\label{figS6}
\end{figure}

The combination of an exponentially suppressed zero-bias resistance (Eq. (\ref{eq:Tdependence})) and the heat balance equation (\ref{eq:HeatBal}) can lead to a thermal bistability, and consequently a voltage jump as a function of the current. In this section, we show that such bistability cannot describe the resistance jump observed in the vicinity of $B_{c2}(0)$ in our samples, nor the sub-critical current resistance that we attribute, instead, to vortex creep. We apply here a reverse analysis similar to that performed in Ref.~\cite{Ovadia09} for the insulating state: We  numerically calculate the $I$-$V$ curves directly from the heat balance equation (\ref{eq:HeatBal}) as well as the $T$-dependence of the zero-bias resistance. The latter is approximated by Eq. (\ref{eq:Tdependence}) that describes  very well the data in the low-$T$ range (see Fig.~\ref{figS2}a). From the high power data in Fig.~\ref{figS4}, we obtain $\Gamma = 0.36\,nW/\mu m^3/K^{5.5}$ for the approximate volume $\Omega = 24\,\mu m^3$. We then solve the heat balance equation for the experimental phonon bath temperatures corresponding to the $B=11.25\,$T measurement and obtain $T_{el}$ as a function of $I$ as shown in Fig.~\ref{figS6}a. The order of magnitude of overheating agrees with the previous analysis in Fig.~\ref{figS3}b for the high $T_{ph}$ curves and at high current bias. The $I$-$V$ curves can be straightforwardly computed from the relation $V =  R(T_{el}) \times I$. The result shown in the semi-logarithmic plots of Fig.~\ref{figS6}b and c leads to three important conclusions. First, the calculated $I$-$V$'s are hysteretic with a switching and retrapping currents whose order of magnitude agrees with the data at $T_{ph}\geq 0.06$~K. Notice that switching (retrapping) current is defined at the transition from the low (high) resistance state to the high (low) resistance state. Second, the retrapping current is constant, independent of $T_{ph}$, whereas it increases on decreasing $T_{ph}$ in the actual experimental data (see Fig. 4a of the main text and Fig.~\ref{figS1}). Third, the switching current increases with decreasing $T_{ph}$ much more rapidly than the data.

The agreement between data and simulations for $T_{ph}\geq 0.06$~K confirms the analysis of the previous section where the heat balance equation was reproduced by the data at high power (Fig.~\ref{figS4}). However, the voltage jumps seen in these simulations, which result from thermal bistability similar to that found in the insulator~\cite{Ovadia09,Altshuler09}, clearly do not match with the resistance jump observed experimentally. One key difference is the constant rettraping current in the simulation (independent of $T_{ph}$), which results from the fact that, when sweeping the current from the resistive state to zero-bias, the electron bath is overheated to a temperature $T_{el}>T_{ph}$ that is independent of $T_{ph}$. In the data of Fig. 4a and Fig.~\ref{figS1}, the retrapping current increases when $T_{ph}$ is lowered, indicating that another mechanism, namely vortex depinning, is in play and drives the resistance jump. Still the small asymmetry between rettraping and switching currents in the data is most likely a residual effect of overheating in the resistive state that decreases the value of the critical current at the retrapping current. At last, the fact that the simulation fails to reproduce the switching current for $T_{ph}< 0.06$~K --the computed switching current diverges much faster that the data-- further demonstrates that the resistance jump does not result from a thermal bistability.

\begin{figure}[h!]
	\includegraphics[width=0.5\linewidth]{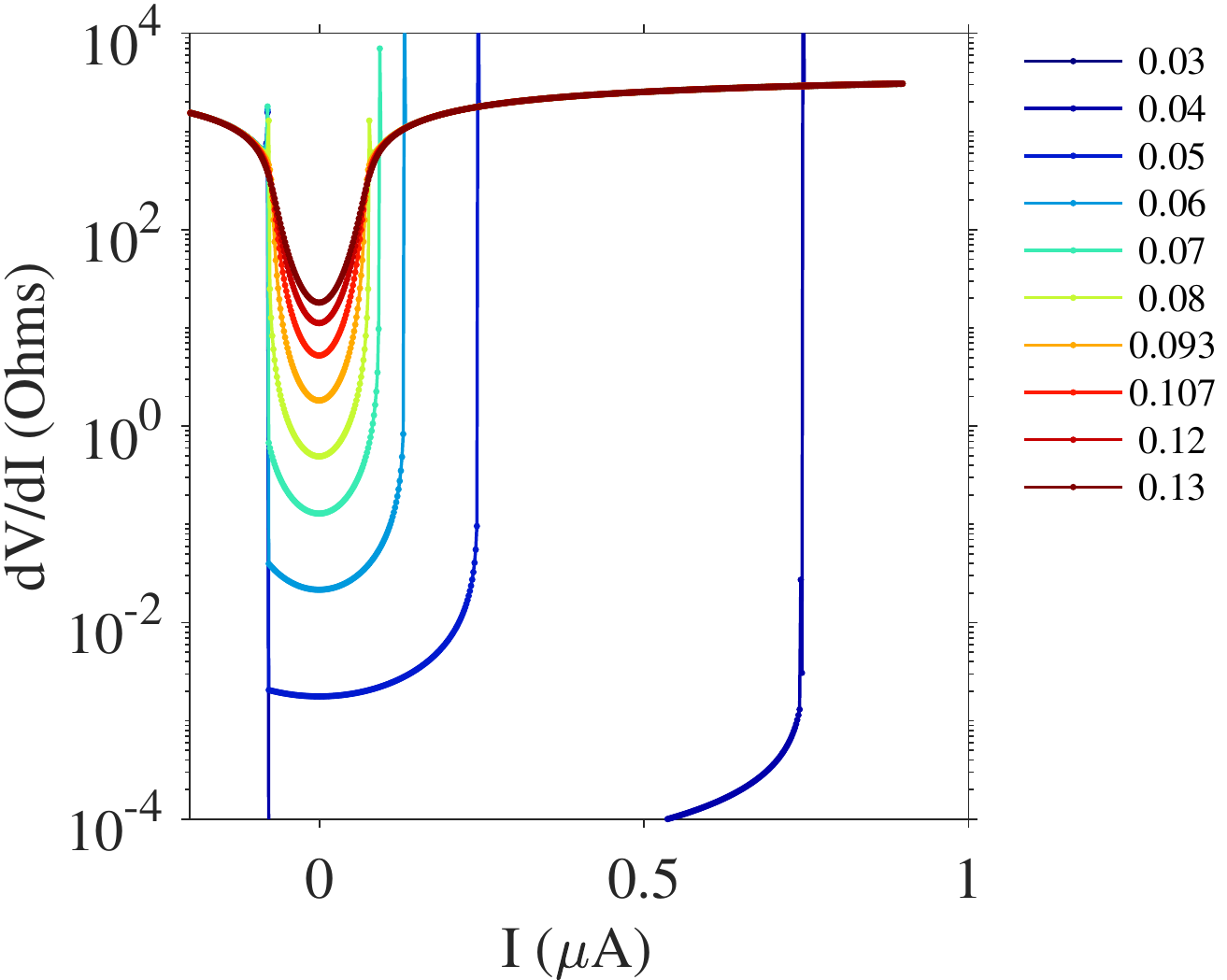}
	\caption{\textbf{Numerically computed differential resistance}.\textbf{a,} $dV/dI$ versus $I$ computed from the $I$-$V$ data of Fig.~\ref{figS6}.}
	\label{figS7}
\end{figure}
We finish our discussion on a last but major difference between the simulated $I$-$V$'s and the data. Below the critical current, an exponential increase of the differential resistance with the applied current is systematically observed in our data (see Fig. 4a and Fig.~\ref{figS1}). We show in Fig.~\ref{figS7} the differential resistance $dV/dI$ numerically computed from the simulations shown in Fig.~\ref{figS6}. We clearly see that, below the switching current, the dependence of the differential resistance on applied current is not exponential at all, marking a stark difference with the data. 

In conclusion, the non-linearities in the $I$-$V$'s in the superconducting state near $B_{c2}(0)$ are intrinsic and results from the vortex glass state. The exponential increase of the resistance below $j_c$ is the signature of vortex creep and the resistance jump signals depinning of the vortex glass. Overheating is still present in the resistive state at large current but does not affect the vortex physics below the critical current. At high bias above the critical current, it is not clear however how to disentangle heating effects with thermal creep interpretation that has been put forward recently~\cite{Thomann12,Buchacek18}. The role of dissipation in this high current regime deserves further theoretical and experimental attention that goes beyond the scope of this work.

\subsection{Magnetic field dependence of the thermal bistability scenario}

\begin{figure}[h!]
	\includegraphics[width=0.4\linewidth]{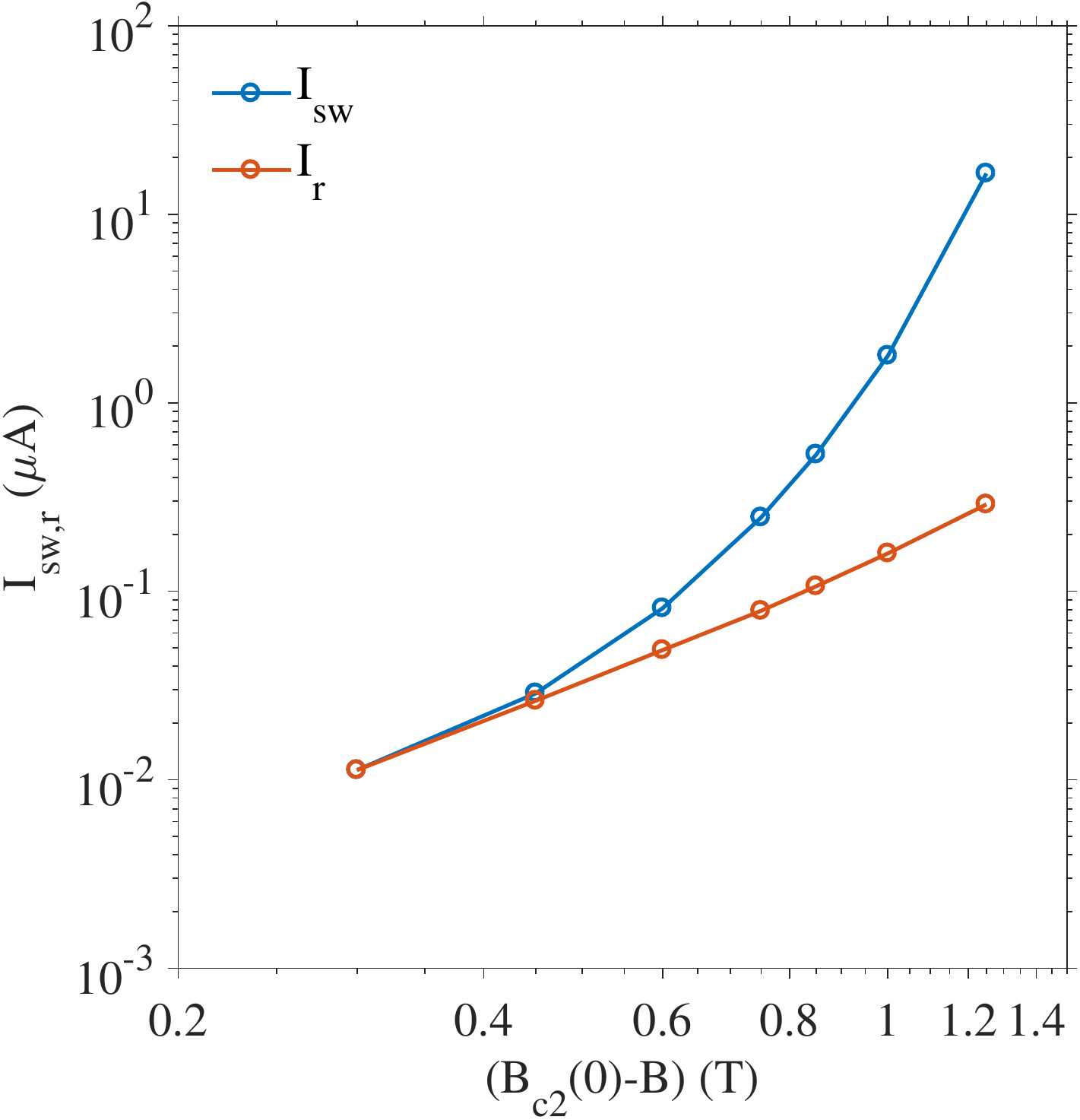}
	\caption{\textbf{Switching and retrapping currents on approaching $B_{c2}(0)$.}. Switching and retrapping currents, $I_{sw}$ and $I_{r}$,  versus  $B_{c2}(0) - B$ computed from the $I$-$V$'s and the $R(T)$ data at various magnetic field values and $T_{ph} = 0.05$~K. Parameters of the heat balance equation are the same as in the previous section.}
	\label{figS8}
\end{figure}
We show in this section the evolution of the switching current with magnetic field $B\approx B_{c2}(0)$ within the heating scenario. For this purpose, we repeat the previous analysis:  We first fit the zero-bias $R(T)$ curves with Eq. (\ref{eq:Tdependence}) for different values of $B$ in the vicinity of $B_{c2}(0)$. This enables us to compute the switching and retrapping currents at a fixed $T_{ph}=0.05\,$K as a function of $B$ and check if the switching and retrapping currents induced by the thermal bistability can mimic the mean-field scaling $j_c \propto (B_{c2}(0) - B)^{3/2}$. The result is shown in Fig.~\ref{figS8} which displays the switching and retrapping currents as a function of $B_{c2}(0) - B$. We see that the rettraping current scales as a power of $B_{c2}(0) - B$ with an exponent $\simeq 2.2$, but the switching current does not. This, again, demonstrates that the thermal bistability interpretation cannot account for the resistance jump nor for the mean-field scaling of the critical current that has been observed in this work.

\section{Derivation of the normalized slope $db(t)/dt\big|_{t\rightarrow0}$.}

Here  we present the transformation between Eq.(15) and Eq.(16) of the main text.
Inserting Eq.(15)  into $b(t)=B_{c2}(T)/(-T_c\, dB_{c2}/dT|_{T=T_c})$ yields
\begin{equation}\label{S1}
\frac{b(0)-b(t)}{t}= \left[1+\sqrt{1+\frac{C_1^2d^2}{\epsilon a_0^2}}\right]\frac{\pi\chi}{24\rho_{s0}d}\frac{B_{c2}(0)}{|dB_{c2}/dT|_{T=T_c}}.
\end{equation}
It remains only to determine $B_{c2}$ in two limits, $T\rightarrow 0$ and $T\rightarrow T_c$, starting with the latter. As discussed in the main text, the transition in the films is of the BKT type with the condition $\rho_s(T,B_c)=\chi{T}/d$. Near $T_c$ the superfluid stiffness of a disordered superconductor is given by~\cite{Tinkham}
\begin{equation}\label{S2}
\rho_s(T,B)=\frac{\pi\sigma_n\hbar}{8Te^2}|\Delta(T,B)|^2.
\end{equation}
The gap $\Delta(T,B)$ near $T_c$ can be found by minimizing the free energy in Eq.(2) of the main text, without the gradient term:  $|\Delta|^2=\alpha/2\beta$. Here $\alpha$ is given by Eq.(3) of the main text,
 which we expand to first order in $B$, and 
 $\beta=\nu\psi^{(2)}(1/2)/32\pi^2T^2$ [$\psi^{(2)}(x)=d^2\psi(x)/dx^2$ is the second derivative of the digamma function].  Thus,  in this limit the superfluid stiffness as a function of temperature and magnetic field is 
\begin{equation}
\label{S3}
\rho_s(T,B)=\frac{2\pi^2{T}\sigma_nh}{e^2\psi^{(2)}(1/2)}\left[\ln\frac{T}{T_{c0}}+\psi'\left(\frac{1}{2}\right)\frac{eBD}{2\pi{c}T}\right].
\end{equation}
Inserting this expression into the BKT condition and taking a derivative with respect to the temperature yields
\begin{equation}\label{S4}
\Big|\frac{dB_{c2}(T)}{dT}\Big|_{T=T_c}=\frac{2\pi\hbar{c}}{eD\psi'(1/2)}. 
\end{equation}
The critical magnetic field at zero temperature, in contrast, is determined by the mean field condition $\alpha=0$
\begin{equation}
\label{S5}
B_{c2}(T=0)=\frac{2\pi\hbar{c}}{eD}T_{c0}e^{\psi(1/2)}.
\end{equation}
Using the expressions in  Eqs.~\eqref{S4} and~\eqref{S5},  we can write Eq~\eqref{S1} as  
\begin{equation}
\label{S6}
\frac{b(0)-b(t)}{t}= \left[1+\sqrt{1+\frac{C_1^2d^2}{\epsilon a_0^2}}\right]\frac{\pi\chi{T_{c0}}}{24\rho_{s0}d}\psi'\left(\frac{1}{2}\right)e^{\psi(1/2)}.
\end{equation}
The final step in estimating $b(t)$ is to find the ratio of
 $\rho_{s0}$ to the mean-field transition temperature $T_{c0}$. 
For moderately disordered superconductors, in the absence
of any pair-breaking mechanism, semiclassical theory yields~\cite{class1,Tinkham}
\begin{equation}
\rho_{s0}= \pi\Delta\hbar\sigma_n/4e^2 =  1.76 \pi\hbar\sigma_nT_{c0}/4e^2.
\label{rhoA}
\end{equation}
We emphasize that the ratio $\rho_{s0}/T_{c0}$ is expected to be reduced for strongly disordered superconductors~\cite{Yazdani,FIKC2010,FI2015}
with respect to the semiclassical  formula in Eq. ~\ref{rhoA}.

For ultra-thin films with $d\ll a_0$ we find, using Eqs.~\eqref{S6} and~\eqref{rhoA},
\begin{equation}\label{S7}
\frac{b(0)-b(t)}{t}= 0.8\frac{e^2}{\hbar\sigma_nd}=0.4\pi\frac{R_{\square}}{R_Q}\equiv K\frac{R_{\square}}{R_Q}.
\end{equation}
Here $R_Q=h/4e^2$ is the quantum of resistance for electron pairs and we used $\chi=2/\pi$, as appropriate for the two-dimensional limit. Since this result relies on Eq.~\ref{rhoA} that overestimates the ratio $\rho_{s0}/T_{c0}$, we expect to experimentally observe larger values for $K$.  For thick films with $d \gg a_0$,  Eqs.~\eqref{S6} and~\eqref{rhoA} imply 
that $|db/dt|$ grows linearly with $R_{\square}/R_{Q}$, i.e.,  
\begin{equation}\label{S8}
\frac{b(0)-b(t)}{t}=\tilde{g}_0 \sqrt{\frac{e^2}{\hbar\sigma_n a_0}} + \tilde{K}\frac{R_{\square}}{R_Q},
\end{equation}
where $\tilde{g}_0 \approx 0.05 \sqrt{C}$, and $\tilde{K} \approx 0.1$.
Note that  $|db/dt|$ does not extrapolate to zero as $R_{\square}=1/\sigma_nd\rightarrow0$.
Similar to the result in the thin-film limit, the true numerical values of both $\tilde{g}_0$ and $\tilde{K}$ are expected to be larger due to the lower value of $\rho_{s0}/T_{c0}$. Moreover, as explained in the main text, the coefficients $\tilde{g}_0$ and $\tilde{K}$ can deviate from the values given by straightforward expansion of Eq.~\eqref{S6} for large $d/a_0$. Such a deviation would reflect higher-order corrections to $\rho_s(T,B)$, which may become important in thick films.

\end{document}